\documentclass{aa}  
\usepackage{txfonts}
\usepackage{graphicx}
\usepackage{txfonts}
\def\hii{\mbox{H\,{\sc ii}}}

\newcommand{\Msol}{\mbox{\rm M$_{\odot}$~}}

\newcommand{\Lsol}{\mbox{\rm L$_{\odot}$~}}

\newcommand{\mic}{\mbox{$\mu$m~}}

\usepackage[]{natbib}

\bibpunct{(}{)}{;}{a}{}{,}
\begin{document} 

\title{Dust and Gas environment of the  young embedded cluster IRAS~18511+0146}

   \author{S. Vig\inst{1}\thanks{Part of this work has been carried out at INAF Osservatorio Astrofisico di Arcetri.}
, L. Testi\inst{2,3}, C. M. Walmsley\inst{2,4},  R. Cesaroni\inst{2}
\and S. Molinari\inst{5}
          }

   \institute{Indian Institute of Space science and Technology, Thiruvananthapuram
 695 547, India \\
              \email{sarita@iist.ac.in}
\and
INAF-Osservatorio Astrofisico de Arcetri, Largo E. Fermi 5, I-50125 Firenze, Italy
\and
European Southern Observatory, Garching, Germany 
\and
Dublin Institute of Advanced Studies, 31 Fitzwilliam Place, Dublin 2, Ireland
\and
INAF-Instituto di Fisica dello Spazio Interplanetario, Via G. Galilei, CP-27, I-00044 Frascati, Italy
            }

   \date{}

\abstract
{Since massive and intermediate mass stars form in clusters, a comparative investigation of the environments of the young embedded cluster members can reveal  significant information about the conditions under which stars form and evolve.}
{IRAS~18511+0146 is  a young embedded (proto)cluster located at 3.5~kpc surrounding what appears to be an intermediate mass protostar. In this paper, we investigate the nature of cluster members (two of which are believed to be the most massive and luminous) using imaging and spectroscopy in the near and mid-infrared. In particular, we examine the three brightest mid-infrared objects and two among these are believed to be most massive ones driving the luminosity of this region.}
{Near-infrared spectroscopy of nine objects (bright in K bands) towards IRAS~18511+0146 has been carried out.
Several cluster members have also been investigated in the mid-infrared using spectroscopic and imaging with VISIR on the VLT.
Far-infrared images from the \textit{Herschel} Hi-GAL survey have been used to construct the column density and temperature maps of the region. }
{The brightest point-like object associated with IRAS~18511+0146 is  referred to as S7 in the present work (designated UGPS J185337.88+015030.5 in the UKIRT Galactic Plane survey). S7 is likely the most luminous object in the cluster as it is bright at all wavelengths ranging from the near-infrared to millimetre. 
Seven of the nine objects show rising spectral energy distributions (SED) in the near-infrared, with four objects showing Br-$\gamma$ 
emission. 
Three members: S7, S10 (also UGPS J185338.37+015015.3) and S11 (also UGPS J185338.72+015013.5) are bright in mid-infrared with diffuse emission being detected in the vicinity of  S11 in PAH bands. Silicate absorption is detected towards these three objects, with an absorption maximum between 9.6 and 9.7~$\mu$m, large optical depths ($1.8 - 3.2$), and profile widths of $1.6-2.1$~$\mu$m. The silicate profiles of S7 and S10 are similar, in contrast to S11 (which has the largest width and optical depth).  
The cold dust emission peaks at S7, with temperature at 26~K and 
column density N(H$_2$)$\sim7\times10^{22}$~cm$^{-2}$. The bolometric luminosity of IRAS~18511 region is L$\sim1.8\times10^4$~L$_\odot$. S7 is the main contributor to the bolometric luminosity, with L (S7) $\gtrsim10^4$~L$_\odot$.}
{S7 is a high mass protostellar object with ionised stellar winds, evident from the correlation between radio and bolometric luminosity as well as the asymmetric Br-$\gamma$ profile. The differences in  silicate profiles of S7 and S11
could be due to different radiation environment as we believe the former to be more massive and in an earlier phase than the latter. } 
 

\titlerunning{Dust and Gas environment of the young embedded cluster IRAS~18511+0146}
\authorrunning{S. Vig et al.}

   \maketitle
\section{Introduction}

\begin {figure*}
\centering
\includegraphics[height=7.0cm]{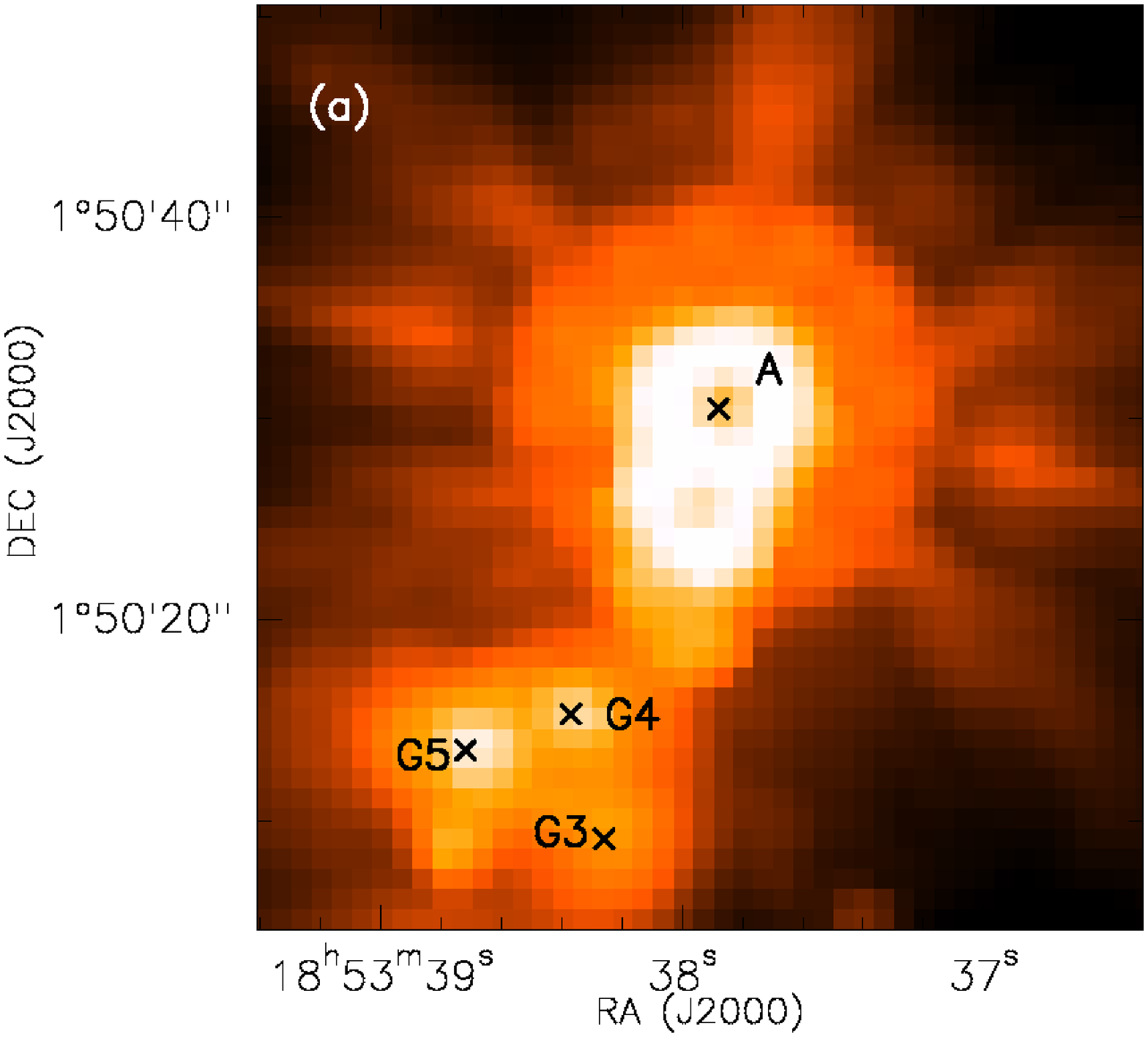}
\hspace*{1cm}
\includegraphics[height=7.0cm]{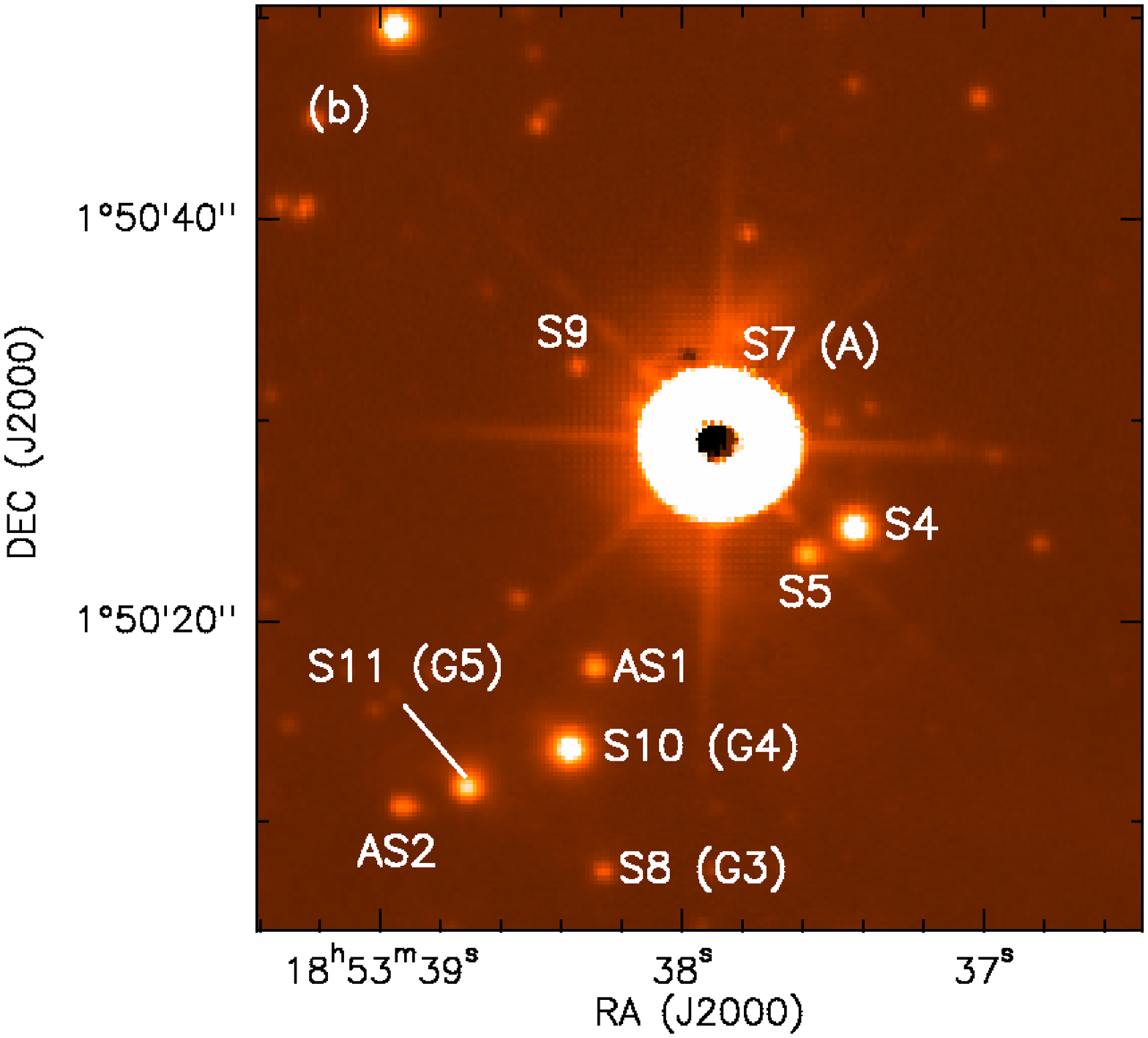}
\caption{(a) \textit{Spitzer}-GLIMPSE 8 $\mu$m and (b) UKDISS - GPS K band image of IRAS~18511 region. (a) The mid-infrared sources 18511-A (S7) as well as young stellar objects G3 - G5 are marked. The sources for which near-infrared spectroscopy have been carried out are shown in (b).}
\label{irac4_ukdiss}
\end {figure*}

It is now widely accepted that massive stars form in clustered environments, deeply embedded in 
molecular clouds \citep{2007ARA&A..45..565M}. In addition, it is found that the distribution of stars in a 
young cluster is related to the distribution of dense gas in the parent 
molecular cloud \citep{2003ARA&A..41...57L}. Therefore, studies of young 
embedded (proto)clusters are useful in understanding the process of clustered 
star formation \citep{2007ARA&A..45..481Z}. In particular, the formation of small stellar groups around 
Herbig Ae/Be stars represents a transition between rich clusters 
harbouring massive stars and loose groups/associations of lower mass stars
\citep{1998A&AS..133...81T}. There are numerous works  investigating star formation using a wide variety of probes that encompass the examination of a single or multiple protostar(s), the embedded cluster, the gas and dust distribution, the dynamics in various stages of evolution, as well as the probable star formation history. However, studies probing clusters of protostars are relatively few, for example \citet{2007AJ....134..346C}, \citet{2012ApJ...745L..30W}, \citet{2003MNRAS.344..809W}. These mostly investigate the massive and rarer end of the initial mass function as other intermediate to low mass cluster members are usually difficult to isolate. This difficulty is not only due to high extinction, but also due to the fact that these infant objects are not radio or mid-infrared loud. With growing interest in trying to understand the conditions of  protocluster formation \citep{2007A&A...472..519A, 2009A&A...505..405P}, surveys are being carried out to locate protoclusters across different regions of the Galaxy \citep{2005ApJS..161..361K}. In this paper, we investigate IRAS~18511+0146 that is a likely forerunner to a Herbig type small star cluster \citep[][hereafter VTW07]{2007A&A...470..977V}. The earlier distance to IRAS~18511 was estimated at 3.9 kpc but a more recent distance estimate from RMS  survey puts the distance to this object as 3.5 kpc while solving the distance ambiguity \citep{2013ApJS..208...11L, 2011A&A...525A.149M}. In the present work, we proceed with the recent distance estimate of 3.5~kpc. The luminosity of this region based on the far-infrared spectral energy distribution is $\sim8.5\times10^3$ \Lsol~\citep{2013ApJS..208...11L}. 
An infrared, submillimetre and radio investigation of this region has revealed the presence
 of a young protocluster in the midst of high extinction filamentary features and clumpy structures (VTW07). The advent of {\it Herschel} has shown that  filamentary structures are ubiquitous towards star forming regions in the Galactic plane, and this is significant as they could undergo collapse themselves to form stars \citep{2010A&A...518L.100M, 2010A&A...518L.102A}.  
In fact, \citet{2009A&A...505..405P} have listed a filamentary structure towards IRAS~18511 in their catalog of {\it Spitzer} dark clouds.  


\begin{table*}
\caption{Details of the sources from UKIRT-GPS, 2MASS and \textit{Spitzer}-IRAC.}
\begin{tabular}{c l c c c c} \hline \hline
S. No &  Source $^a$ & UKIRT - GPS$^b$ &H & K$_s$ & 8.0 $\mu$m \\
   & Id. & Designation &(mag) & (mag) & (mag)   \\ \hline
1 &S4                  & UGPS J185337.42+015026.2 & $11.34\pm0.01 $ &  $11.06\pm0.01$  & ... \\ 
2 &  S5                & UGPS J185337.58+015024.9 &$13.24\pm0.01$ &  $12.81\pm0.01$ & ... \\ 
3 & S7$^{c}$ (A) & UGPS J185337.88+015030.5 &$9.30\pm0.03$ &  $6.61\pm0.03$ & ...  \\ 
4 & S8 (G3)         & UGPS J185338.26+015009.1 & $17.40\pm0.04$ & $14.38\pm0.01$ &  $6.09\pm0.27$ \\ 
5 &  S9                 & UGPS J185338.34+015034.3 & $15.03\pm0.01 $ &  $14.24\pm0.01$ & ... \\ 
6 & S10 (G4)       & UGPS J185338.37+015015.3 & $15.57\pm0.01 $ &  $11.17\pm0.01$ &  $4.95\pm0.07$ \\ 
7 &  S11 (G5)       & UGPS J185338.72+015013.5 &$ 18.03\pm0.07$ &  $12.32\pm0.01$ &   $4.32\pm0.04$ \\ \hline \hline  
\multicolumn{6}{c}{Additional Sources} \\ 
\hline 
8 & AS1                &UGPS J185338.29+015019.3 & $15.97\pm0.01 $ & $13.32\pm0.01$ & ... \\ 
9 & AS2                & UGPS J185338.91+015012.4 & $ 15.84\pm0.01$ & $13.81\pm0.01$ & ...\\ 
 \hline \hline
\end{tabular}
 \label{sour}
\vskip 0.2cm
$^a$ Nomenclature from VTW07. AS\# represents additional sources.\\
$^b$ Designation includes coordinates in J2000 as  JHHMMSS.ss+DDMMSS.s \\
$^c$ Saturated in the UGPS H and K band images as well as in the IRAC bands. The magnitudes given for H and K bands are from 2MASS.  \\
\end{table*}


Using near and mid-infrared colours, we identified 16 young stellar objects in this region in our earlier study (VTW07). We had also carried out cluster simulations of a young embedded cluster whose results are consistent with a cluster comprising Class I and Class II sources. 
The presence of high extinction knots and filamentary structures towards this region has been explained as arising due to compression by stellar winds from cluster members based on simulations by 
\citet{2008ApJ...684.1384R}. That the cluster members are highly embedded is evident from the high extinction values obtained from the submillimetre maps (VTW07) with the peak visual extinction towards IRAS~18511 being $\sim86$ magitudes. The extinction towards this region has also been explored by \citet{2012A&A...537A..27G} using the 3.4~\mic absorption feature (characteristic of diffuse ISM) and they cite association with a remnant diffuse cloud as one of the reasons for explaining the exceptionally high values of optical depth of this feature.    

Our main objective in the present study is to examine the nature of the cluster members in IRAS~18511. The near and mid-infrared emission towards the cluster is shown in Fig.~\ref{irac4_ukdiss}. More than two-thirds 
of the luminosity of this region is driven by the brightest object here, 2MASS18533788+0150305 or IRAS~18511-A (designated S7 following VTW07). Some of the cluster members identified by VTW07 are also shown in the figure. The selection of cluster members in VTW07 was based on colour-colour diagrams (using near and mid-infrared wavebands). In order to confirm the nature of the objects and whether they are a part of the young embedded cluster, we have carried out near-infrared spectroscopic observations of objects bright in the K-band. In addition, 
K band spectroscopy can be used to probe the presence of ionised and  shocked gas in the immediate vicinity of the young stellar objects through the Br-$\gamma$ and H$_2$ lines. We also carried out ground based high resolution imaging of this region in the mid-infrared for the following reasons: (i) to probe the nature of extended morphology of S7 observed in the IRAC~8~\mic image which is saturated, shown in Fig.~\ref{irac4_ukdiss} (a), and (ii) to resolve other bright mid-infrared sources that might be present in this region. Further, we examine the extinction due to silicates towards the cluster members  bright in mid-infrared. In addition, we have explored the far-infrared emission of this region using images from the Hi-GAL survey of the \textit{Herschel Space Observatory}. In this paper, we present the results of these observations and  examine few individual cluster members in detail. In Sect.~2, we 
outline the observations while the results are presented in Sect.~3. These are 
discussed in detail in Sections 4 and 5. Finally, a summary of  our results is presented in Sect.~6.

\section{Observations and Data reduction}

The near and mid-infrared observations of IRAS~18511 were carried out using the ESO Telescopes at the La Silla and Paranal Observatories under the programme ID 079.C-0166.

\subsection{Near-infrared spectroscopy using NTT-SOFI}

The objects in the region associated with IRAS~18511+0146 for which near-infrared spectroscopic observations were carried out using the instrument SOFI \citep[Son OF Isaac;][]{1998Msngr..91....9M}
are shown in the K band image of the UKDISS Galactic Plane Survey \citep[UGPS;][]{{2007MNRAS.379.1599L},{2008MNRAS.391..136L}}  in Fig.~\ref{irac4_ukdiss}~(b). The details of these sources are given in Table~\ref{sour}. While seven sources are selected from the list of young stellar objects compiled in VTW07 (Table~\ref{sour}), the 
spectra of two additional sources in the vicinity were also taken.  These sources are denoted by AS1 and AS2.
The H and K magnitudes of the targets listed in the table are taken from the UGPS Archive \citep{2008MNRAS.384..637H}.

SOFI is an infrared spectrometer located at the Nasmyth A focus of the ESO 3.6 m New Technology 
Telescope (NTT) located at La Silla, Chile. The observations were carried out during the nights of 23 and 25 July 
2007. The red grism with a slit of width $1''$ and pixel scale 0\farcs292 per pixel was used. This achieves a slit length 
of 4\farcm8. We used the low spectral resolution mode, with a resolving 
power of $R\sim 600$, covering a wavelength range $1.53-2.52$~$\mu$m. Taking 
advantage of the long slit, the spectra of nine 
objects were taken using six slit positions.  The exposure times ranged from 200 to 480 seconds depending on the magnitude of the source. The typical seeing was less than 
1\arcsec~except for the case of  S7 where the seeing was closer to 2\arcsec. 
In order to remove the effects of the telluric absorption features from the spectra, a star with a similar airmass was observed as the telluric standard. We observed Hip089677 (spectral type B5V) and Hip101505 (spectral type B3IV) during 23 and 25 July 2007, respectively.

The data reduction was carried out using the IRAF package. In brief, the bad 
pixels were corrected followed by flat-fielding of all the images (source as well as 
standard star) to correct for any artifacts or distortions caused by pixel-to-pixel variation 
in sensitivity. The remaining sky background has been removed by subtracting one of  
the (paired) dithered frames from the other. The wavelength 
solution from the arc image (xenon lamp), which was also corrected for pixel-to-pixel variations, was applied to the images. The source as well as standard star spectra were then extracted using the task 
`apextract' in IRAF. Features intrinsic to the standard star were identified 
using \citet{1998ApJ...508..397M} and \citet{1997ApJS..111..445W} and removed 
from the source spectra by linearly interpolating over them. The telluric 
features from the standard star spectrum were aligned with the source
spectra by cross-correlating them. The source spectra were then divided by the 
standard star spectrum followed by multiplication of a black-body curve with 
the $T_{\rm eff}$ corresponding to the standard star. Finally, the flux levels 
were scaled to achieve the best possible agreement between the spectral data 
and the H and K$_s$ photometric fluxes of the object 
after correcting for the response functions of the latter in the two bands.

\subsection{Mid-infrared imaging and Spectroscopy using VLT-VISIR}

We have carried out mid-infrared imaging and spectroscopic observations of 
the IRAS~18511+0146 region using the  ESO-VLT imager and 
spectrograph VISIR \citep{2004Msngr.117...12L}, mounted 
on the Cassegrain focus of the VLT Unit Telescope 3 (Melipal). The observations were 
carried out on the nights of 26 and 27 July 2007. 

\subsubsection{Imaging}

The imaging was carried out with a scale of 0\farcs127 per pixel leading to a 
field-of-view of 32\farcs3$\times$32\farcs3. Two adjacent fields covering S7 and S11 respectively,
were imaged in various bands. The imaging of these fields were carried out using various N band filters such as ArIII 
(8.99 $\mu$m), SiV\_1 (9.82 $\mu$m) and SiV (10.49 $\mu$m), as well as the Q1 band (17.65 $\mu$m). 
In addition, the field covering S11 was also imaged using the PAH1 
(8.59 $\mu$m), PAH2 (11.25 $\mu$m), PAH2\_2 (11.88 $\mu$m) and Q3 
(19.5 $\mu$m) filters. The optical seeing was better than 0\farcs8 throughout the 
observations. The airmass at the time of observations varied in the range $1.1-2.0$. A number of standard stars were 
observed for calibration in order to maintain conditions of similar airmass through filters for both the fields. These include 
HD198048, HD78788, HD22663, HD16815, HD187150, HD18695, and HD199642. Chopping as well as nodding were carried out in order to minimise the background. The integration time ranged between 15 - 40 minutes depending on the source and filters used.

The VISIR data reduction pipeline provided by ESO was used to reduce the data.
This involves coadding the elementary images to 
obtain chopping-corrected data to remove the high mid-infrared background. 
The residual background was finally removed by combining the nodding 
positions to create the final image.  The full-width at half maxima (FWHM) in the 
different wavelength bands are $\sim$0\farcs32 - 0\farcs35 in the N band and $\sim$0\farcs5 in the Q band.

\subsubsection{Spectroscopy}

The spectroscopy of S7, S10 and S11 were carried out in N-band during the night of  27 July 2007 using a slit of size 0\farcs75$\times$32\farcs3. 
All but one spectroscopic observations were 
carried out in the low resolution mode ($R\sim350$) within the spectral range 7.7 -- 13.5 $\mu$m.
Several standard stars were observed through the gratings under conditions of 
similar airmass as the science targets. These include HD188154, HD199642 and HD4815. The integration time ranged between 10 - 35 minutes depending on the source being observed. 
The brightest object, S7, was also observed using the high resolution ($R\sim25000$) grating centered on the [\ion{Ne}{ii}] 
line at 12.81 $\mu$m. Here the integration time was 60 min. For the high resolution observations, calibration was carried out by observing the asteroid Hygieas. In all the observations, chopping and nodding were carried out to minimise the background. 

The VISIR spectroscopic pipeline provided by ESO was used for data reduction, where the images are first corrected for chopping and nodding followed by correction using a reference frame of the 
infrared background. The next step involves the extraction of the spectrum. 
This is carried out by summing over the pixels within the line profile along 
the spatial axis for the positive as well as the negative beams that are 
present. The values corresponding to the negative beams are subtracted from 
the positive beams. The spectrum thus obtained is then calibrated 
in wavelength using the atmospheric lines from the background frame and any
 wavelength shift, if present, is corrected for. The spectrum is divided 
by the corresponding spectrum of the standard star observed under similar 
conditions of air-mass to remove the effect of the telluric features. The 
flux calibration is carried out by multiplying this resultant spectrum with 
the modelled spectrum of the standard star.

\subsection{Far-infrared images using Herschel Hi-GAL survey}

We have also used far-infrared images from the Herschel Infrared Galactic Plane Survey  \citep[Hi-GAL;][]{2010PASP..122..314M} carried out by the \textit{Herschel Space Observatory}. The \textit{Herschel Space Observatory}\footnote{Herschel is an ESA space observatory with science instruments provided by European-led Principal Investigator consortia and with important participation from NASA.} carrying a 3.5-m passively cooled telescope was launched in May 2009 by ESA \citep{2010A&A...518L...1P}. It consists of three instruments: Photodetector Array Camera and Spectrometer (PACS), Spectral and Photometric Imaging REceiver (SPIRE) and 
Heterodyne Instrument for the Far Infrared (HIFI). 

The Hi-GAL survey used the two instruments: PACS \citep{2010A&A...518L...2P} and SPIRE \citep{2010A&A...518L...3G} in parallel mode, to carried out a survey of the inner Galaxy ($|l|\le60^\circ$, $|b|\le1^\circ$) in five bands: 70, 160, 250, 350 and 500~$\mu$m. The reduced and calibrated images provided by the Hi-GAL team have been used here. A flux calibration uncertainity of 15\% is considered \citep{2010A&A...518L..92W}.  The resolution of the SPIRE images are 18\farcs5, 25\farcs3 and 36\farcs9, at 250, 350 and 500~$\mu$m bands, respectively. The point spread functions of the PACS images at 70 and 160~$\mu$m are 5\farcs8$\times$12\farcs1 and 11\farcs4$\times$13\farcs4, respectively. The scan speed and direction decides the elongation in the images \citep{2016A&A...591A.149M}.
The pixel sizes are 3\farcs2, 4\farcs5, 6\farcs0, 8\farcs0 and 11\farcs5, at 70, 160, 250, 350 and 500 $\mu$m, respectively. These Hi-GAL images have been used to investigate the flux density distribution and to construct the temperature and column density maps of the region associated with the protocluster. 

\section{Results}

\subsection{Near-infrared identification of cluster members}

\begin {figure}
\resizebox{\hsize}{!}{\includegraphics[height=20.0cm]{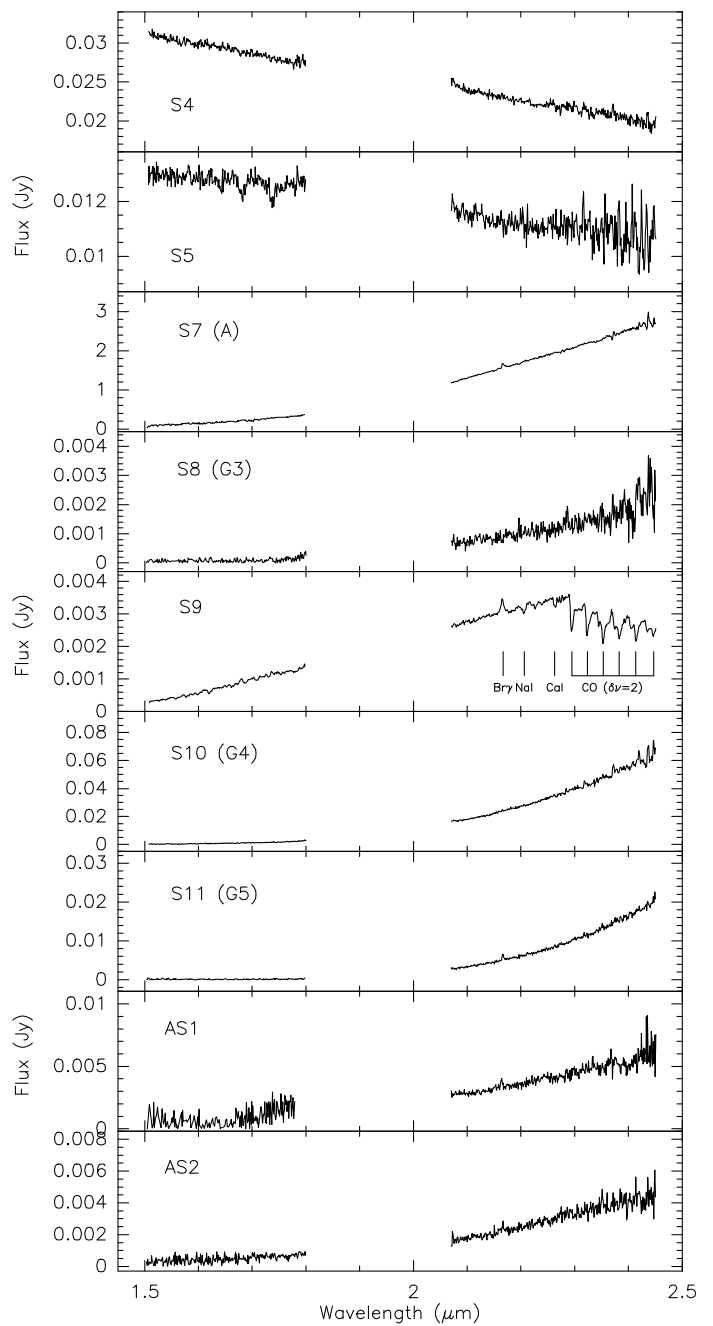}}
\caption{Near infrared HK spectra of objects in the region associated with IRAS~18511. The gap in the spectra between $1.8 - 2.1$~$\mu$m is due to atmospheric absorption.}
\label{nir_spec}
\end {figure}

\begin {figure}
\resizebox{\hsize}{!}{\includegraphics[height=9.0cm]{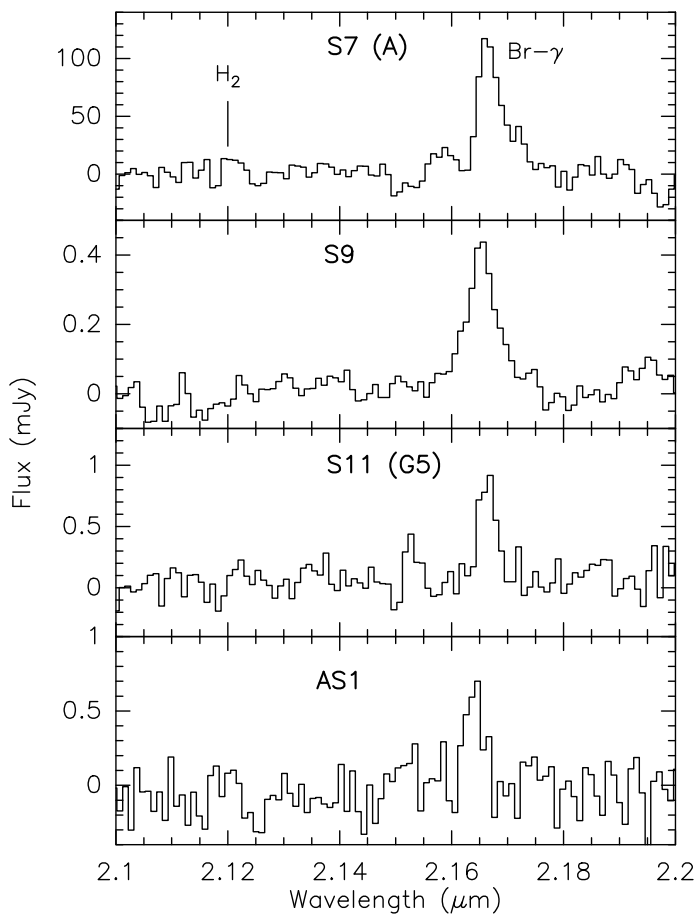}}
\caption{Continuum subtracted spectra of S7, S9, S11 and AS1 showing Br-$\gamma$ line emission. The location of H$_2$ (1-0) S(1) line is also shown. 
}
\label{brgam}
\end {figure}

The near-infrared HK spectra of the sources in the IRAS~18511 region are shown in Fig.~\ref{nir_spec}. The gap in the spectra between 1.8 and $\sim 2.05$~$\mu$m is due to the presence of strong telluric water absorption. The sources S4 and S5 show decreasing continua, while all the other sources have fluxes increasing with wavelength.  
The Br-$\gamma$ line is seen in emission towards S7, S9 and S11 and there is a marginal detection towards AS1.
A magnified view of the wavelength range around the Br-$\gamma$ emission towards these sources can be seen in Fig.~\ref{brgam}. S9  appears to have broad red and blue wings in the Br-$\gamma$ emission profile while S7 shows hints of broadening towards the red edge. The ro-vibrational H$_2$ S(1-0) line at 2.12 $\mu$m, associated with shocked gas from outflows, is not detected in any spectrum. A number of absorption lines are seen in the spectrum of S9 (discussed below).  

S7, the brightest member of the cluster, is believed to be a very young intermediate mass/massive pre-main sequence star (VTW07). The spectrum of S7 is featureless with the exception of Br-$\gamma$ line emission. The flux in the Br-$\gamma$ line is $\sim3.6\times10^{-13}$~erg~cm$^{-2}$~s$^{-1}$. 
The 3-$\sigma$ upper limits to the H$_2$ emission corresponds to $1.1\times10^{-13}$~erg~~cm$^{-2}$~s$^{-1}$. 
 S8, S10 and S11 are classified as Class I sources based on mid-infrared colours and show positive slopes of the SED between near and mid-infrared wavebands (VTW07).  Except for Br-$\gamma$ emission towards S11, all the spectra are fairly featureless. The absence of any absorption lines in the spectra of these sources is strikingly similar to that of low mass Class I young stellar objects, whose featureless spectra are explained on the basis of high continuum veiling due to the surrounding circumstellar emission by \citet{2000AJ....120..430G}.

The sources S4 and S5 are located close to each other, with a separation of 2\farcs5. They have low near-infrared excesses (J-H $\sim0.3-0.5$; H-K $\sim0.5 - 0.8$) and are not detected in mid-infrared (Spitzer-IRAC) images. We searched the optical images to locate their optical counterpart(s) and found an optical source in the Digital Sky Survey (DSS) red and blue images. However, the resolution of DSS images is relatively poor ($\sim5''$) and it is not possible to distinguish if either or both emit in optical. However, both are quite similar and their spectra do not show any features either in emission or absorption.  Although these sources are located close to S7 in projection (within $8''$), the extinction towards the latter is high as it is invisible in optical images.  S4 and S5, on the other hand, suffer low extinction as they are visible in optical and therefore, they are likely to be late type foreground objects. The absence of absorption lines can be explained by the veiling due to circumstellar dust that gives rise to the infrared excess. 

AS1 and AS2 are sources that were not considered for the classification of young stellar objects in VTW07 as they were not detected in the Palomar J band image but were detected in H and K bands. Their locations as well as infrared colours (H-K~$\sim2.6$ and 2.1, respectively) suggest that they are likely to be young stellar objects. They are, however, not detected in the mid-infrared images (Spitzer-IRAC and VISIR), implying lower flux densities and consequently, low mass young stellar objects. Their spectra also do not show indications of any absorption or emission lines, other than the marginal detection of Br-$\gamma$ emission from AS1 at the 2$\sigma$ level (Fig.~\ref{brgam}).

\begin {figure*}
\includegraphics[height=7.0cm,angle=0]{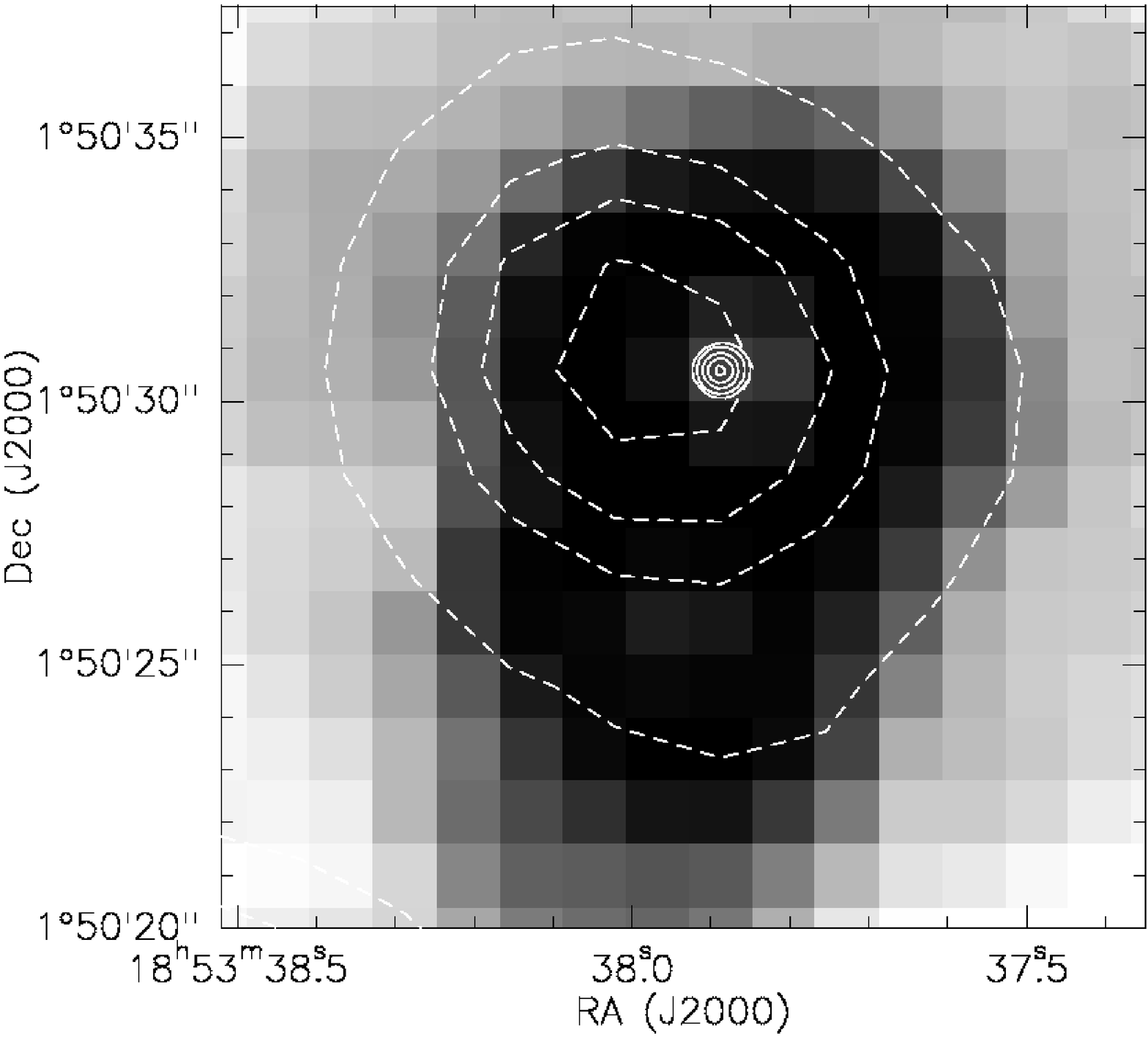}
\hspace*{1cm}
\includegraphics[height=7.0cm,angle=0]{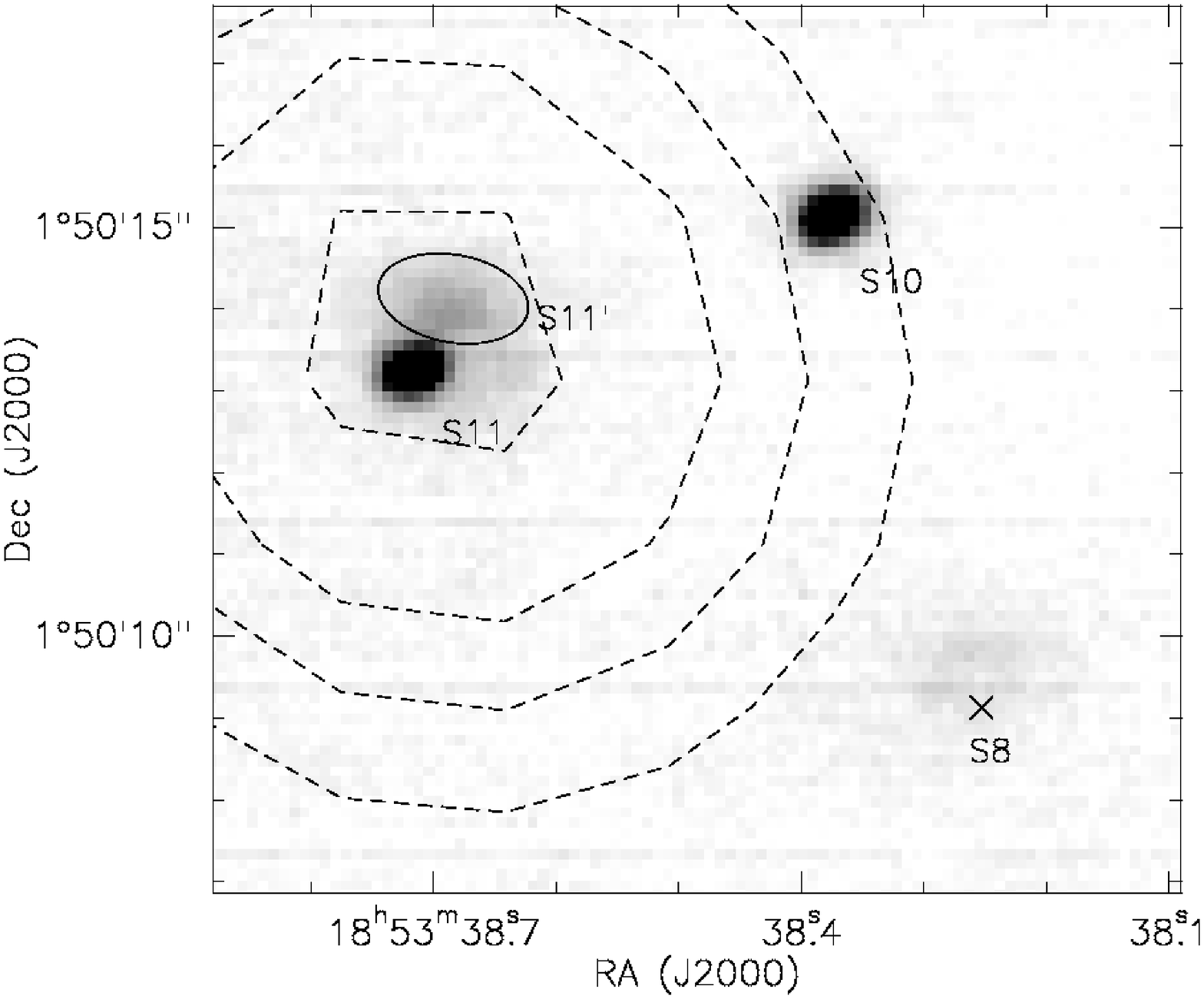}
\caption{(Left) The IRAC image of S7 at 8.0~$\mu$m with the 17.7~$\mu$m emission shown as solid line contours.  (Right) The image shows S10 and S11 through 8.6~$\mu$m (PAH1) filter. An ellipse is drawn to show the location of S$11'$, diffuse emission associated with S11. The dotted contours in both the panels represent the radio emission at 8.5 GHz  at levels 0.04, .08,  0.1 and 0.12 mJy/beam (left) and 0.2, 0.3, 0.4 and 0.55 mJy/beam (right) where beam size is 9\farcs2$\times$7\farcs5 \citep{2011ApJ...739L...9S}.
}
\label{mir}
\end {figure*}

S9 is unique among our sample as it displays a number of absorption lines other than the Br-$\gamma$ emission line. These include the atomic blended  lines of [\ion{Na}{i}] doublet (2.206 and 2.209~$\mu$m), [\ion{Ca}{i}] triplet (2.261, 2.263 and 2.266~$\mu$m) and molecular CO absorption bandheads. The CO bandheads at 2.239 ($\nu$=2-0), 2.323 (3-1), 2.353 (4-2),
 2.383 (5-3), 2.414 (6-4), 2.446~\mic (7-5)~corresponding to $\delta\nu=2 $ are detected. However, the [\ion{Fe}{ii}] line at 1.643~$\mu$m characteristic of shocked gas is not detected. The stronger red continuum would seem to imply that this is a young stellar object as the atomic lines and molecular  bandheads in absorption, as well as Br--$\gamma$ emission are also seen in spectra of low mass Class II and flat spectrum objects \citep{1996AJ....112.2184G}. However, a closer inspection of the spectrum shows that the $^{13}$CO absorption bandheads are also resolved, and other blended lines such as [\ion{Fe}{i}] lines at 2.227 and 2.239~\mic~\citep{2000AJ....120.2089F} are also detected. These two features ($^{13}$CO bandheads and Fe lines in absorption) are seen towards late-type evolved stars  such as KM giants and dwarfs \citep{1997ApJS..111..445W}. But these objects usually do not show  Br-$\gamma$ in emission. If Br-$\gamma$ is present, it is in weak absorption.
The equivalent widths of the absorption lines are relatively large:  [\ion{Na}{i}] - 5.6\AA, [\ion{Ca}{i}] - 4.3\AA, CO($\nu$=0-2) - 
14.2\AA~and Br-$\gamma$ in emission: 8.4\AA. 
If we do not consider the Br-$\gamma$ emission, the spectrum can be said to convincingly belong to a late spectral type evolved object for the following reasons: (i) The clear detection of $^{13}$CO counterparts along with $^{12}$CO bandheads also favours evolved later type objects \citep{2009A&A...494..253K}. (ii) The presence of atomic Fe~I absorption lines is seen only towards  evolved late type objects. (iii) The atomic and molecular absorption features are very strong with relatively large  equivalent widths that are seen towards lower surface gravity objects \citep{2004A&A...425..489C}. Nevertheless, it is important to note that a few Class III young stellar objects with large equivalent widths in the absorption lines of  [\ion{Na}{i}],  [\ion{Ca}{i}] and CO have been observed such as Hubble 4, and HBC~360 among the sample of \citet{1996AJ....112.2184G}.
Although all this strongly suggests that S9 is an evolved late type spectral object, we cannot ignore (i) the presence of Br-$\gamma$ emission whose equivalent width is 8.4~\AA, 
and (ii) a rising red continuum. The latter can be explained by the presence of large extinction towards S9. 
We also compared S9 with the spectra of FU-Orionis type objects that show features in their post-eruption phase similar to late type objects \citep{2002AJ....124.2194R}. Further, FU Orionis-like stars have not been observed to clearly show Br-$\gamma$ emission. In this regard, \citet{2010AJ....140.1214C} speculate that the relationship between Br-$\gamma$ emission line flux and mass accretion collapses at very high mass accretion rates. Hence, there is a possibility that S9 could be a post-outburst object with the spectrum showing features of transition to that of a later spectral type.
Since Br-$\gamma$ emission has been detected from nearly $50$\% of low mass young stellar objects \citep{1996AJ....112.2184G} and 75\% of massive young stellar objects \citep{2013MNRAS.430.1125C}, its presence in S9 compels us to consider other possibilities that can explain all the features. It is possible that S9 is a late evolved type star that shows variability in Br-$\gamma$ emission similar to HR7564~$\chi$Cyg \citep{1997ApJS..111..445W}, although it is weak in the latter object. Alternately, there is a remote possibility of two objects (young stellar object and evolved object) being projected towards the same line-of-sight. 
Finally, one cannot rule both objects being part of a binary system.

It is interesting to note that if we assume that all objects other than S4, S5 and S9 are young, embedded and belong to the cluster, then we have been able to detect Br-$\gamma$ from three out of six sources.
Our sample is obviously too small to derive reliable statistics. Nonetheless, our finding that about half of the sample shows detectable Br-$\gamma$ emission, is consistent with other studies of young stellar objects in groups  \citep{1996AJ....112.2184G,2013MNRAS.430.1125C}.

\subsection{Cluster members in Mid-infrared}

\begin{table}
\caption{Photometric mid-infrared fluxes of sources associated with IRAS~18511 
observed using VLT-VISIR. The errors on the fluxes are of the order of 
$\sim15$\% mainly due to calibration errors of standard stars.}
\begin{tabular}{c c c c c c} \hline \hline
Filter & Wavelength & \multicolumn{4}{c}{Source Flux (Jy)} \\ 
\cline{3-6}
 & band ($\mu$m) & S7 & S10 & S11 & S$11'$ \\ \hline
 & & & & & \\
PAH1 & 8.6 & -     & 0.56 & 0.62 & 0.17 \\
ArIII & 9.0 & 11.5  & 0.52 & 0.33 & -    \\
SiV\_1 & 9.8 & 9.8   & 0.31 & 0.13 & -    \\
SiV & 10.5 & 10.6 & 0.27 & 0.14 & -    \\
PAH2 & 11.3 & -    & 0.57 & 0.36 & 0.10 \\
PAH2\_2 & 11.9 & -    & 0.72 & 0.44 & -    \\
Q1 & 17.7 & 28.2 & 1.04 & 0.35 & 0.22 \\
Q3 & 19.5 & -    & 1.06 & 0.39 & 0.31 \\
\hline
\end{tabular}
\label{fluxes}
\end{table}

Mid-infrared emission from a young stellar object arises when small dust grains in its vicinity (e. g. from disk or envelope) are heated by far-UV photons and the spatial extent of the emission is determined by the optical depth of these dust grains.  
Our ground based mid-infrared images show compact (unresolved) emission from S7, S10 and S11.  
 The emission from the brightest infrared object in this region, S7 at 17.7~$\mu$m is shown as solid line contours in Fig.~\ref{mir} (left) while the emission through the PAH1 filter at 8.6~$\mu$m from the field covering S8, S10 and S11  is shown in 
Fig.~\ref{mir} (right). A comparison with IRAC images from \textit{Spitzer}-GLIMPSE in VTW07 indicates that the source S8 which lies 
in the field of S10 and S11, is barely detected at 8.6~$\mu$m in Fig.~\ref{mir} (right).

Figure~\ref{mir} (left) shows the emission from S7 through the 17.7~$\mu$m filter band
overlaid on the IRAC~8~$\mu$m image. We perceive that the high resolution VLT-VISIR image of S7 is unresolved unlike the IRAC 8.0~$\mu$m image that shows extended emission and is saturated. This suggests that  
the extension seen in the \textit{Spitzer}-IRAC 5.8 and 8.0~$\mu$m images are probably artifacts. The emission from S10 and S11 at mid-infrared wavelengths in Fig.~\ref{mir} (right) shows 
the presence of diffuse emission in the vicinity ($\sim$0\farcs8) of S11 towards the north as well as west of S11, with stronger emission 
towards the north-west. We call this emission knot $\rm{S}11'$. This emission (size 
$\sim 2''\times1''$) is 
particularly strong in the PAH filters: 8.6~$\mu$m (PAH1) and 11.3~$\mu$m (PAH2). 
Further, this emission from  $\rm{S}11'$ is stronger at the longer wavelength Q bands at 17.7 and 19.3~$\mu$m, 
implying a rising spectral energy distribution.

The flux densities of the sources S7 and S10 have been extracted by using aperture 
photometry with an aperture radius of 1\arcsec$\;$ and a background annulus 
of thickness 1\arcsec$\;$located at distance of 4\arcsec$\;$from the centre 
of the aperture. This method cannot be used to extract the flux of S11 due to the presence of 
$\rm{S}11'$ in the vicinity.  
For S11, we fitted a gaussian distribution along with a constant 
function (corresponding to the background) and extracted the flux by 
integrating the emission under this distribution after subtracting the 
background emission. To estimate the flux densities from  $\rm{S}11'$, we used a smaller aperture of diameter 
$\sim0$\farcs8 and the background has been estimated using an aperture of 
similar size in the vicinity where there is no detectable emission. The 
calibration of fluxes has been carried out using the standard stars. The
error in photometry is $\sim15$\%, which is mainly due to the uncertainity in the fluxes
of standard stars as well as due to airmass corrections. The flux densities
of S7, S10, S11 and $\rm{S}11'$ are listed in Table~\ref{fluxes}.

\begin {figure}
\resizebox{\hsize}{!}{\includegraphics[bb=100 120 450 700]{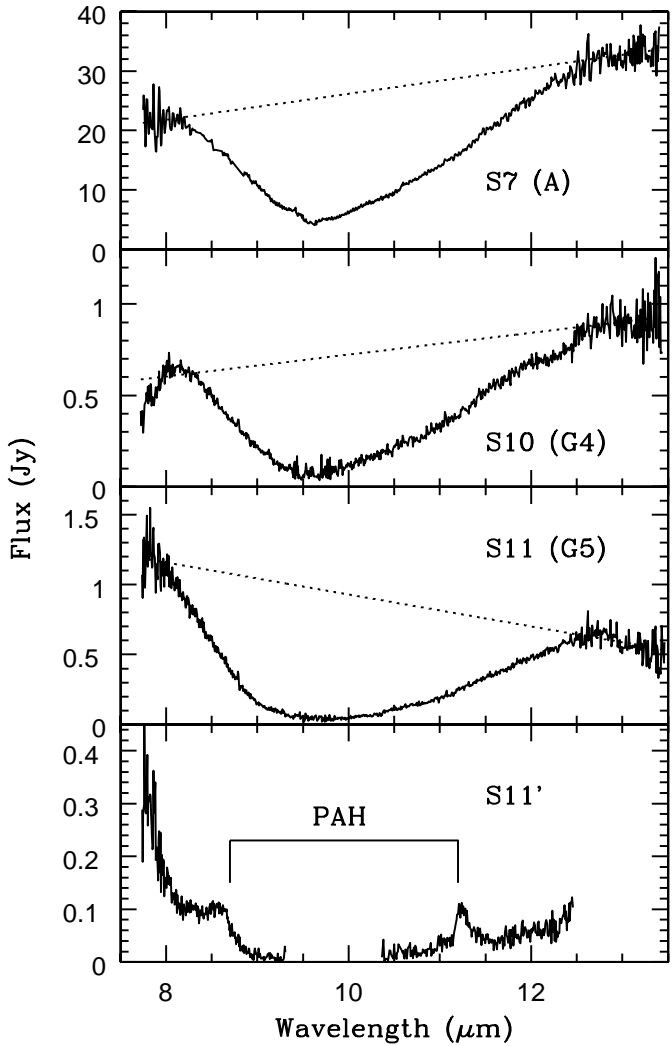}}
\caption{
Mid-infrared spectra of bright cluster members in the IRAS~18511 region along with the spectrum of S$11'$. The dotted line in each panel represents the fitted continuum. 
}
\label{mir_spec}
\end {figure}

$\rm{S}11'$ is not detected at near-infrared wavelengths, and has been detected solely in 
the PAH bands as well as the longer wavelength 17.7 and 19.3~$\mu$m bands (see Table~\ref{fluxes}). This is evident from its mid-infrared spectrum shown in Fig.~\ref{mir_spec}.  $\rm{S}11'$ could  either be (i) a faint embedded object (that has bright PAH features) enveloped in diffuse emission,
or (ii) a diffuse emission knot associated with S11. 
It is difficult to deduce the nature of $\rm{S}11'$ from the present mid-infrared images alone and necessitates higher sensitivity images.

The mid-infrared spectra of S7, S10, S11 and $\rm{S}11'$ 
are shown in Fig.~\ref{mir_spec}. 
The spectrum for $\rm{S}11'$ has discontinuities near $\sim9.8$ and 
$\sim13$ $\mu$m as the emission is weak, and noise dominates. The spectra of all the sources show  
silicate absorbtion features. Closer inspection reveals that the silicate profile features appear different for S7, S10 and S11 and this is discussed later, in Section 4. 
The Polycyclic 
Aromatic Hydrocarbons (PAHs) emission at 8.6 $\mu$m and 11.2 $\mu$m is seen only towards $\rm{S}11'$.
The flux in the PAH1  band is larger than that in PAH2 band as the measured fluxes are $5.6\times10^{-13}$ and $3.4\times10^{-13}$~erg~s$^{-1}$~cm$^{-2}$ in the 8.6 and 11.2 $\mu$m bands, respectively.

In order to obtain the optical depth profiles of these sources, we fitted and subtracted a linear function
to the continua represented by flux densities below 8~\mic and longer than 12.7~$\mu$m. The differences are negligible  ($< 5$\%) when the continuum is considered as a second order polynomial. 
The continua of S7 and S10 show an increase with wavelength while the continuum of S11
shows a negative power law slope. The optical depth profiles are estimated using the expression corresponding to the solution of the radiative transfer equation in the absence of emission.

\begin{equation}
I_{\rm obs}(\lambda) = I_{\rm o}(\lambda) \exp^{-\tau_{\lambda}}
\end{equation}
\begin{equation}
\tau_{\lambda} = \ln\left(\frac{I_{\rm o}(\lambda)}{I_{\rm obs}(\lambda)}\right) 
\end{equation}

\begin {figure}
\resizebox{\hsize}{!}{\includegraphics[bb=30 160 545 676]{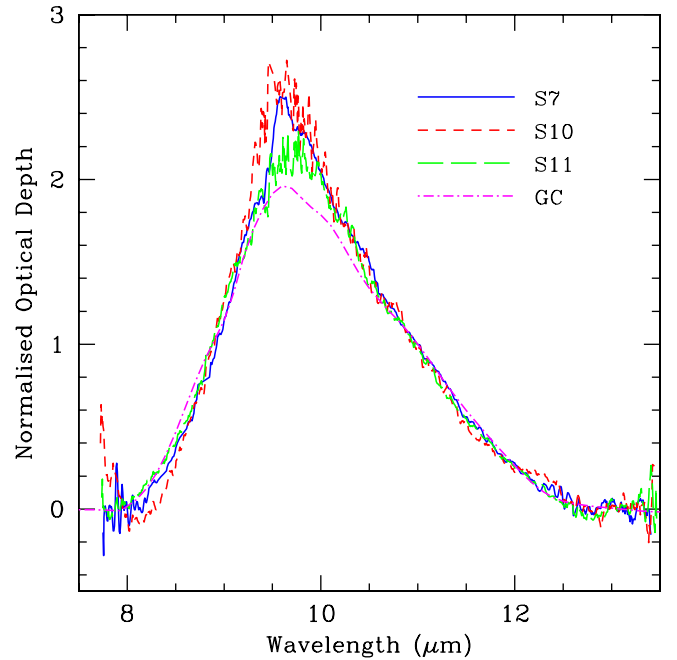}}
\caption{
Opacity profiles of the silicate feature of the cluster members (solid line, short-dashed line, long-dashed line) normalised at 11~$\mu$m. Also shown is the silicate profiles of the Galactic centre (dot-dashed line) from \citet{2006ApJ...637..774C}.
}
\label{sil_norm}
\end {figure}

Here $I_{\rm obs}$ is the observed flux at a given wavelength within the absorption profile and $I_{\rm o}(\lambda)=I_{\rm cont}(\lambda)$ represents the fitted continuum. The properties of the silicate profiles 
including the peak optical depth are listed in Table~\ref{sil_tab}. The optical depths peak at wavelength $\sim9.6-9.7$~\mic~and 
the maximum optical depth values range between $1.8-3.2$.
We have normalised the optical depth profiles at the longward end of the absorption, at 11~$\mu$m \citep{2011A&A...526A.152V}.
The normalised optical depth profiles are shown in Fig.~\ref{sil_norm} and discussed in detail in the next section. The FWHM of the profiles are $\sim1.6$ and $2.0$~$\mu$m for S7 and S11, respectively.

\begin{table*}
\begin{center}
\caption{Details of silicate absorption features and extinction towards cluster members.}
\begin{tabular}{c c c c c c c} \hline \hline
Object & $\lambda_{\rm p}^a$ & $\Delta \lambda^b$ & $\tau_{\rm max}^c$ & A$_{\rm V}^d$ &  A$_{\rm V-ISM}^e$ & A$_{\rm V-mc}^f$  \\ 
 &   ($\mu$m) & ($\mu$m) & &  (mag)  & (mag) & (mag)\\ \hline
 & & & & & &\\
S7 & 9.6 & 1.6 &  $1.8\pm0.2$ & $35\pm8$ & $16.3\pm3.0$ & $19\pm9$\\
S10 & 9.6 & 1.6 &  $2.2\pm0.2$ & $43\pm9$ & $16.0\pm4.3$ & $27\pm10$ \\
S11 & 9.7 & 2.0 &  $3.2\pm0.5$ & $62\pm17$ & $17.5\pm5.0$ & $45\pm18$ \\
\hline
\end{tabular}
\label{sil_tab}
\end{center}
\vskip 0.05cm
 \scriptsize{$^a\lambda_{\rm p}$ - Wavelength at which the silicate absorption is maximum. \\
$^b\Delta \lambda$  - Width of silicate profile based on wavelengths at which absorption is half the peak value.  \\
$^c\tau_{\rm max}$ - Maximum optical depth of the silicate profile. \\
$^d$A$_{\rm V}$ - Visual extinction estimates obtained using Eqn. (6). \\
$^e$A$_{\rm V-ISM}$ - Visual extinction estimates associated with diffuse interstellar medium based on the 3.4~$\mu$m aliphatic hydrocarbon feature by \citet{2012A&A...537A..27G}. \\
 $^f$A$_{\rm V-mc}$ - Contribution of dense molecular cloud to visual extinction based on values obtained from A$_{\rm V}$ and A$_{\rm V-ISM}$}\\

\end{table*}

The mid-infrared spectra do not show evidence of the [\ion{Ne}{ii}] line at 12.81 
$\mu$m from any of the studied objects. Although radio continuum emission and Br-$\gamma$ line emission have been detected towards
S7 and S11, we have not detected [\ion{Ne}{ii}] towards them.
Our aim in carrying out the high-resolution spectroscopy of S7 at 
12.8 $\mu$m was to search for the presence of the [\ion{Ne}{ii}] line. This 
high resolution spectrum is shown in Fig.~\ref{highres}. It allows us to 
put an upper limit of $3\times10^{-14}$~erg~s$^{-1}$~cm$^{-2}$ on the 
[\ion{Ne}{ii}] line from S7. The upper limits obtained 
for S10 and S11 are  $7\times10^{-14}$ and 
$4\times10^{-14}$~erg~s$^{-1}$~cm$^{-2}$, respectively.

\subsection{Far-infrared emission}

All the Hi-GAL emission maps show the brightest emission towards S7. At 70 and 160~$\mu$m, we see two emission peaks. The brighter one corresponds to S7 while the secondary peak corresponds to S10 and S11. At longer wavelengths, we see only one peak as the latter are not resolved. The Hi-GAL images are shown in Appendix A.

\subsubsection{Temperature and Column Density Maps}

Since the cluster is embedded in the molecular cloud, it is instructive to obtain the temperature 
and column density distribution of this region. It is possible to construct these maps on the scale of a pixel using the five far-infrared Hi-GAL wavelengths, JCMT SCUBA maps at 450 and 850~$\mu$m  (VTW07), ATLASGAL\footnote{The ATLASGAL project is a
collaboration between the Max-Planck-Gesellschaft, the European Southern
Observatory (ESO), and the Universidad de Chile.} \citep{{2010Msngr.141...20S},{2014A&A...565A..75C}}, and SEST-SIMBA \citep{2006A&A...447..221B}. This would involve regridding and convolving all the images to the lowest resolution of 36\farcs9 corresponding to the Hi-GAL 500~$\mu$m image. As the long wavelength regime  ($>100$~$\mu$m) is well sampled, we prefer to construct higher angular resolution images (18\farcs5) of column density and dust temperature using fewer far-infrared images.  For this, we use the Hi-GAL images at 70, 160 and 250~$\mu$m as well as the JCMT-SCUBA images at 450 and 850~$\mu$m. As the pixel sizes and resolutions are different for these
images, we first regridded and convolved the images to the largest pixel size 
($6''\times6''$) and lowest resolution (18\farcs5) corresponding to the 
250~$\mu$m image using the Herschel Interactive Processing Environment (HIPE)\footnote{HIPE is a joint development by the Herschel Science Ground Segment Consortium, consisting of ESA, the NASA Herschel Science Center, and the HIFI, PACS and SPIRE consortia.}. The kernels used for convolution of the PACS and SPIRE images are those given by \citet{2011PASP..123.1218A}. For the SCUBA images, we used Gaussian kernels of size $10''$ and 15\farcs5 at 450 and 850~$\mu$m, respectively. In this way, we obtained images at five wavelengths between 70 and 850~$\mu$m with identical resolution, pixel size and pixel orientation. For every pixel, a spectral energy distribution 
was constructed \citep{2010A&A...518L..92W,2012MNRAS.426..402F} using 
the flux density values ($F_\nu$) and fitted with a modified blackbody spectrum of form:

\begin {figure}
\resizebox{\hsize}{!}{\includegraphics[height=5.0cm]{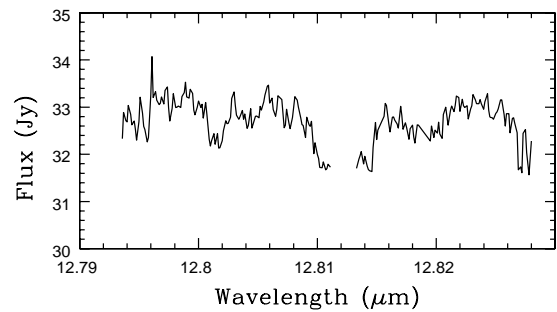}}
\caption{High resolution mid-infrared spectrum of S7 near
the [\ion{Ne}{ii}] line. 
}
\label{highres}
\end {figure}

\begin{equation}
F_\nu = \Omega\,B_\nu(T_D)(1-e^{-\tau_\nu}) 
\end{equation}

\begin{equation}
  \tau_\nu = \mu\,m_H\,k_\nu\,\rm{N(H_2)}
\end{equation}

Here, $\Omega$ corresponds to the solid angle of the pixel, $T_D$ is the dust temperature,
$\tau_{\nu}$ is the optical depth, $m_H$ is the mass of hydrogen atom, $\mu$ is the mean
molecular weight in units of $m_H$, $k_\nu$ is the dust opacity and $\rm{N(H_2)}$ is the molecular hydrogen column density. 
For a gas-to-dust mass ratio of 100, the dust opacity was assumed to vary with frequency as
follows \citep{1983QJRAS..24..267H,1990AJ.....99..924B,2010A&A...518L.102A}.

\begin{equation}
k_\nu = 0.1 \times \left( \frac{\nu}{1000\, \rm{GHz}}\right)^\beta\,\rm{cm}^{-2}\,\rm{g}^{-1}
\end{equation}

Here $\beta$ is the dust emissivity index. Recently, there has been a lot of work in trying to understand factors that contribute to $\beta$ and its variation towards star forming regions, molecular clouds, as well as the diffuse interstellar medium \citep{{2003A&A...404L..11D},{2010A&A...520L...8P}}. Considering that our region of interest is relatively small ($\sim2$ arcmin$^2$), we neglect any change in the value of $\beta$ and assume a constant value of 2 found towards many star forming regions \citep{{2002MNRAS.329..257W},{2013ApJ...767..126S}}. The best fits were obtained using nonlinear least squares Marquardt-Levenberg algorithm, considering $\rm{N(H_2)}$ and $T_D$ as free parameters. Based on other studies, we assumed flux uncertainties of the order $\sim15$\% in all the bands \citep{2013A&A...551A..98L}.

\begin {figure}
\includegraphics[height=6.0cm]{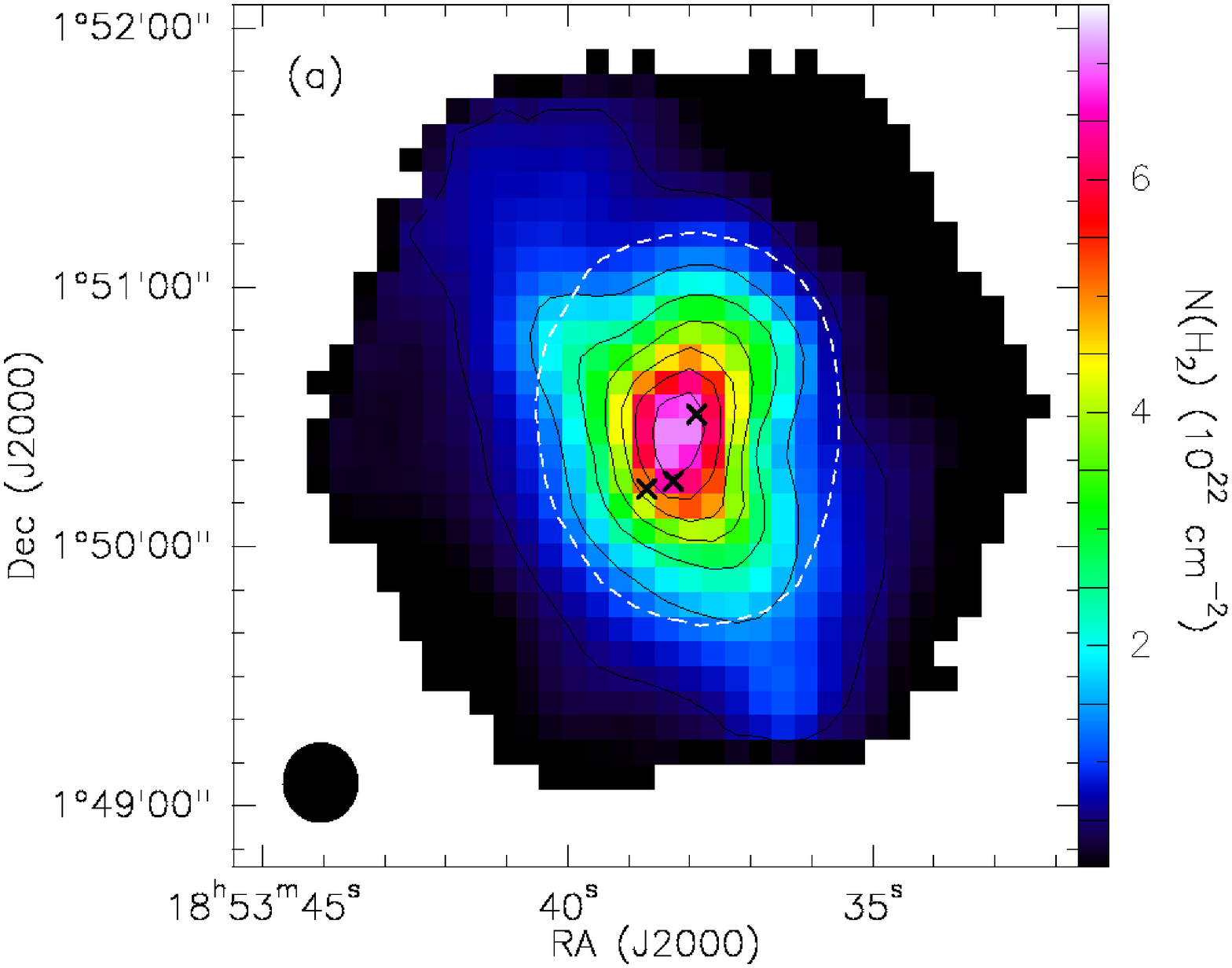}
\includegraphics[height=6.0cm]{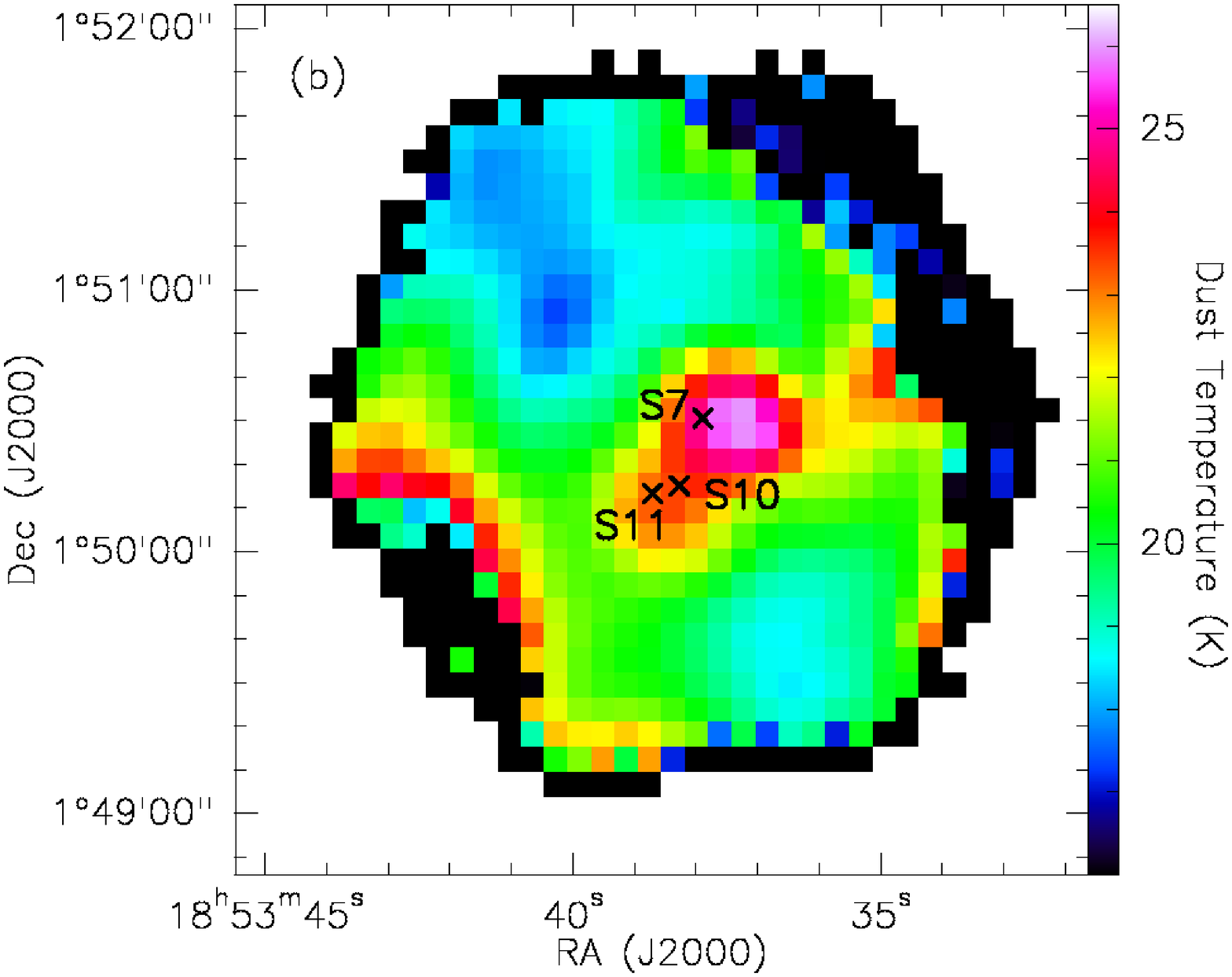}
\includegraphics[height=6.0cm]{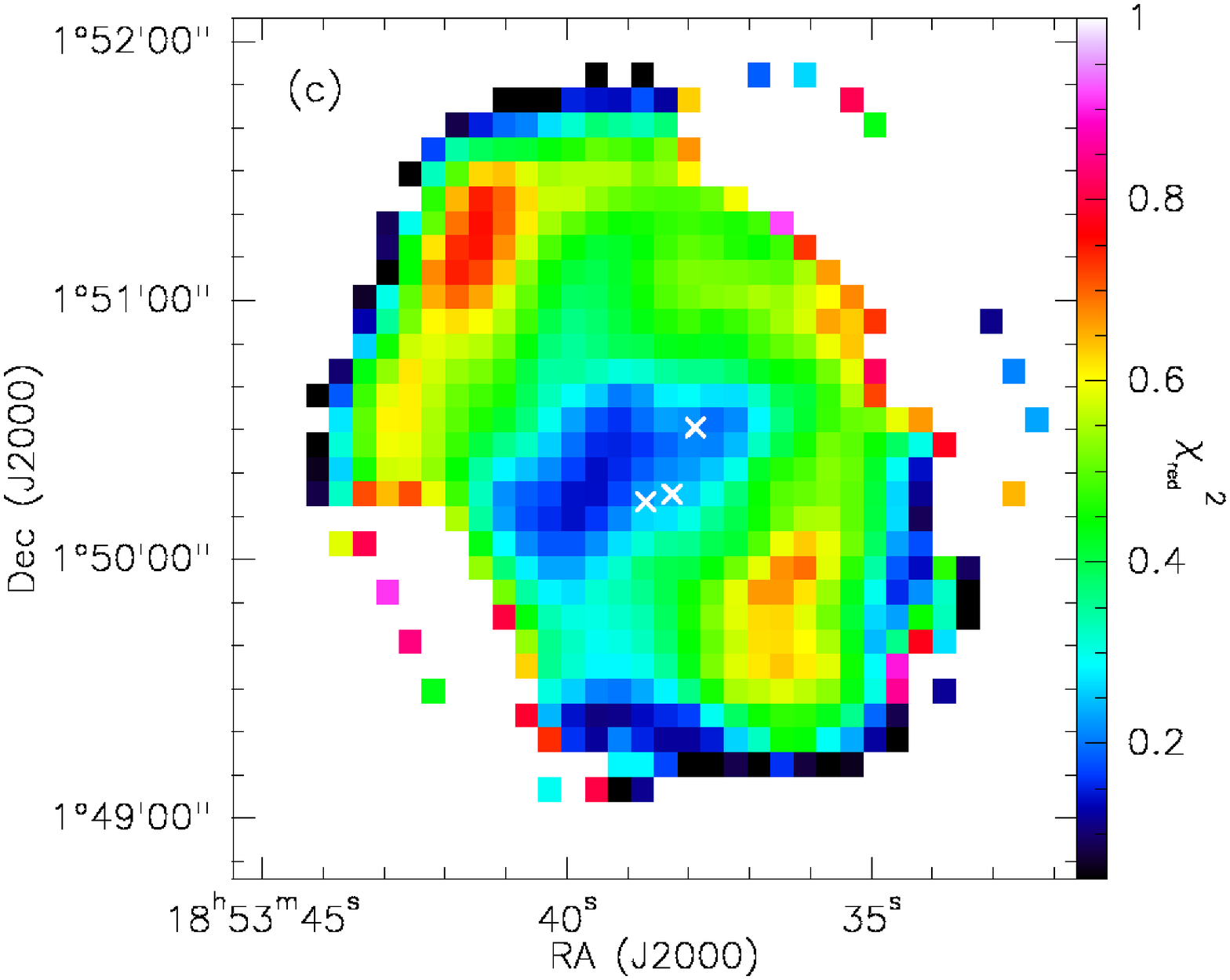}
\caption{ (a) Column density map of $\rm N(H_2)$, (b) Dust temperature map, and (c) $\chi^2_{red}$ map obtained using pixel-by-pixel modified blackbody fit to 70, 160, 250, 450 and 850~$\mu$m fluxes from the maps of IRAS~18511+0146. The beamsize is shown as a filled circle in the lower left corner of (a). The region enclosed within the dashed line in (a) represents the aperture for estimating the SED of this region. The crosses in all the panels depict the positions of sources S7, S10 and S11, shown in (b). }
\label{higal}
\end {figure}

\begin {figure}
\resizebox{\hsize}{!}{\includegraphics[height=5.5cm, angle=-90]{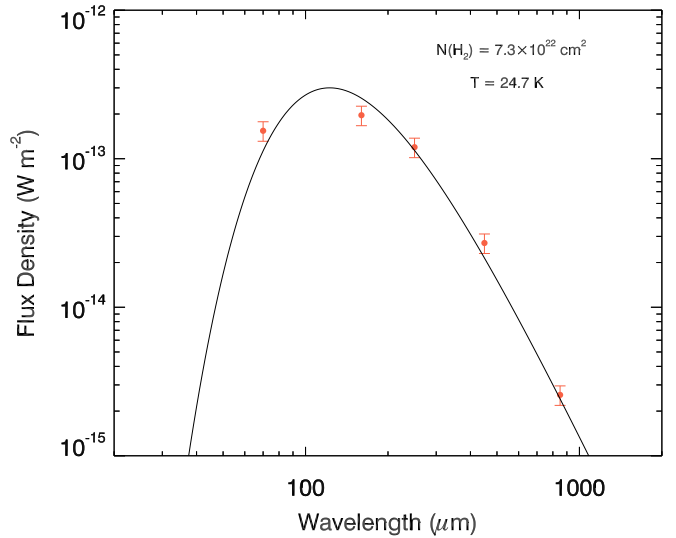}}
\caption{ Modified blackbody fit to the flux densities corresponding to the pixel with maximum column density (J2000 coordinates 18$^h$53$^m$38.08$^s$~+01$^\circ$50$'$30.5$''$). }
\label{bb}
\end {figure}

 The column density, dust temperature and reduced chi-square ($\chi^2_{red}$) maps  are shown in Fig.~\ref{higal}. Both the column density and temperature peak near S7. For further analysis, we consider a region within the 35\% 
contour level of the peak emission at 500~$\mu$m towards IRAS~18511. This excludes the contribution of a 
neighbouring source C (VTW07). An additional reason is that the emission from filamentary 
structures in the vicinity makes it difficult to disentangle the effect of filamentary and background emission from 
the cloud. This region (encircled within the dashed line over the column density image shown in Fig.~\ref{higal}) is also used to estimate the bolometric flux described later, in Section 5.  For all pixels within this region, the modified blackbody fits are quite good as 
 $\chi^2_{red}< 0.7$. The blackbody fit corresponding to the pixel with the maximum column density is shown in Fig.~\ref{bb}. The temperature values range between $18.6 - 26.2$~K with errors between 0.5 and 1.9 K.  The column density values in this region ($1.6\times1.2$~pc$^2$) range from 
$\rm{N(H_2)}\sim0.3-7.3\times10^{22}\,\rm{cm}^{-2}$ with errors ranging between $13-23\%$.  
By integrating the column densities within this region, the total mass for this region is estimated to be $920\pm17$~\Msol. Assuming this mass to be distributed in a spherical volume of radius 0.7~pc, we obtain the mean density as ${\rm n_H}=2.8\times10^4$~cm$^{-3}$.
The mass estimate determined is consistent with the estimates of $750-1310$~\Msol obtained by VTW07 based on SCUBA 
850~\mic emission for the same dust emissivity coefficient, but larger than 559~\Msol estimated by \citet{2006A&A...447..221B} who use an opacity of $\kappa=1$~cm$^2$g$^{-1}$ at  $1.2$~mm wavelength. 
Using C$^{18}$O emission, \citet{1999ApJS..125..143W} estimated the mass to be 826~M$_\odot$. 
The low $\rm{N(H_2)}$ column density structures tracing
cold dust filamentary emission extending north-east and south-west are visible in blue in the column density and temperature maps. The column density follows the
 longer wavelength fluxes closely.

In the dust temperature distribution, apart from the hottest region near S7, we observe a high temperature shoulder extending south-east towards S10 and S11. The high temperature regions indicate active star formation. In addition, we see low temperature region ($15-18$~K) towards the north-east and south-west. These correspond to the filamentary emission seen in the column density map.

\section{Silicate Absorption}

In this section, we discuss results obtained using the mid-infrared spectra towards S7, S10 and S11. Silicate absorption as a 
measure of extinction and the nature of dust along the lines-of-sight of these three objects are discussed. 

\subsection{Measure of Extinction}

For dust in diffuse interstellar medium, the 
optical depth at 9.7 $\mu$m obtained from the silicate absorption feature is 
known to correlate with extinction \citep{1998ApJ...497..824M} as
\begin{equation}
 \frac{\rm{A}_V}{\tau_{9.7}}=19.3\pm2.2
\end{equation}
The extinction values obtained using this relation are listed in Table~\ref{sil_tab}. 
The visual extinction values of the cluster members lie between $35-62$ mag implying that these are deeply embedded objects. 
However, it is important to realise that (i) the above A$_V$ relation does not hold for all diffuse lines-of-sight, and 
(ii) an extrapolation of this relation from diffuse 
lines-of-sight to molecular clouds is not reasonable as the types of silicates encountered there are likely to be different.
\citet{2007ApJ...666L..73C} have investigated the silicate feature in dense clouds and they  
show that the optical depth does not increase monotonically with extinction. These authors find that with a
few exceptions, $\tau_{9.7}$ falls well below the diffuse ISM correlation line. This would imply that the 
extinction values are actually larger than the values obtained here. Thus, according to the relations obtained by \citet{2007ApJ...666L..73C}, the estimates of visual extinction listed in Table~\ref{sil_tab} can be considered as lower limits. 

Irrespective of the magnitude of extinction, what is reliable is the relative extinction between 
objects within the same molecular cloud.  Our results indicate that among the luminous young stellar objects / protostars, the 
most deeply embedded object appears to be S11, followed by S10 and then S7.  This is consistent with the optical depth values obtained for the  3.1~\mic water 
ice feature detected towards these three objects \citep{2012A&A...537A..27G}. The contribution of diffuse ISM towards 
these three lines-of-sight, investigated by the same authors using the 3.4~\mic aliphatic hydrocarbon feature, is found to be nearly constant. If we  assume a ratio of $\rm{A}_{\rm V-ISM}/\tau_{3.4}=250$, this corresponds to  
an interstellar extinction of $\rm{A}_{\rm V-ISM}\sim16-18$~mag. This can be used to estimate lower limits to extinction due to the 
molecular cloud and circumstellar extinction, i.e $\rm{A}_{V-mc}$. In our case, $\rm{A}_{V-mc}$  would be in the range $20-45$~mag. The values of  $\rm{A}_{V-ISM}$  
and $\rm{A}_{V-mc}$ are tabulated in Table~\ref{sil_tab}.

\subsection{Nature of Dust}

The dust along the line-of-sight to an embedded young stellar object comprises contributions from the diffuse interstellar medium, the dense 
molecular cloud as well as the circumstellar environment, that could also include high extinction 
regions such as an accretion disk. In this section, we investigate the silicate features towards
 the cluster members S7, S10 and S11 and compare them relative to each other as well as with 
the silicate profiles observed towards other regions. 

The optical depth profiles of the cluster members are shown in  Fig.~\ref{sil_norm} with the profiles peaking near the wavelength
$9.6-9.7$~$\mu$m.  A relative comparison shows that the profiles towards S7 and S10 (Table~\ref{sil_tab}) are similar whereas
there are a few differences towards S11. This is also evident from Fig.~\ref{mir_spec}. This would seem to imply 
either a change in the circumstellar environment or an effect of disk inclination towards these objects. 
S10 is closer to S11 in projected distance (separation of $5''\sim$ 0.1~pc), while the luminous object S7 is located at a projected distance nearly three times further away (projected separation $17''\sim$ 0.3~pc). Thus, 
the circumstellar conditions can play a vital role in the appearance of the silicate feature even within the same molecular cloud. A 
closer inspection would necessitate knowledge of the nature of dust along the lines-of-sight. 

The shape and peak of the silicate absorption feature is decided by the size, composition, shape and type (amorphous / crystalline) of dust grains. The peak absorption at 9.6~\mic is consistent with that seen around massive protostars \citep{1999A&A...349..267D}. The 
models used for describing the dust towards such massive embedded objects invoke the presence of  amorphous 
pyroxenes to explain the shift in peak towards shorter wavelengths from 9.7~\mic~found towards 
diffuse ISM. For example, \citet{2004ApJ...609..826K} modelled the dust towards the Galactic centre as having nearly 
15\% amorphous pyroxenes and 85\% amorphous olivines by mass for spherical submicron sized grains. The 
width of the silicate profiles is narrower towards S7 and S10, $\Delta\lambda\sim1.6$~$\mu$m, while it is broader towards S11, $\Delta\lambda\sim2.0$~$\mu$m. These widths are relatively 
narrow when compared to the diffuse ISM as well as massive protostars.  The widths of features towards massive protostars 
are broad ($\sim2.3-3.5$~\mic), an indication of coagulated grains that manifests itself as porosity or  
chaotic effects of processing \citep{1994A&A...292..641J, 1999A&A...349..267D}. In fact, \citet{2003MNRAS.340.1173B}
found a bimodal distribution when they plotted the peak absorption wavelength versus the width (FWHM). For sources with absorption peaking at 
 9.6~$\mu$m,~ they found the widths are relatively broad $3.4$~$\mu$m, while the narrower profiles ($\sim2.6$~$\mu$m) peak at 9.8~$\mu$m.
\citet{2005ApJ...622..404K} investigated the shape and peaks of the silicate absorption feature for a 
number of embedded young stellar objects. They find the features peak at the constant wavelength of $9.6\pm0.2$ $\mu$m 
while the FWHM varies between 1.8 and 3.2 $\mu$m. 

\begin {figure}
\resizebox{\hsize}{!}{\includegraphics[bb=75 140 501 680]{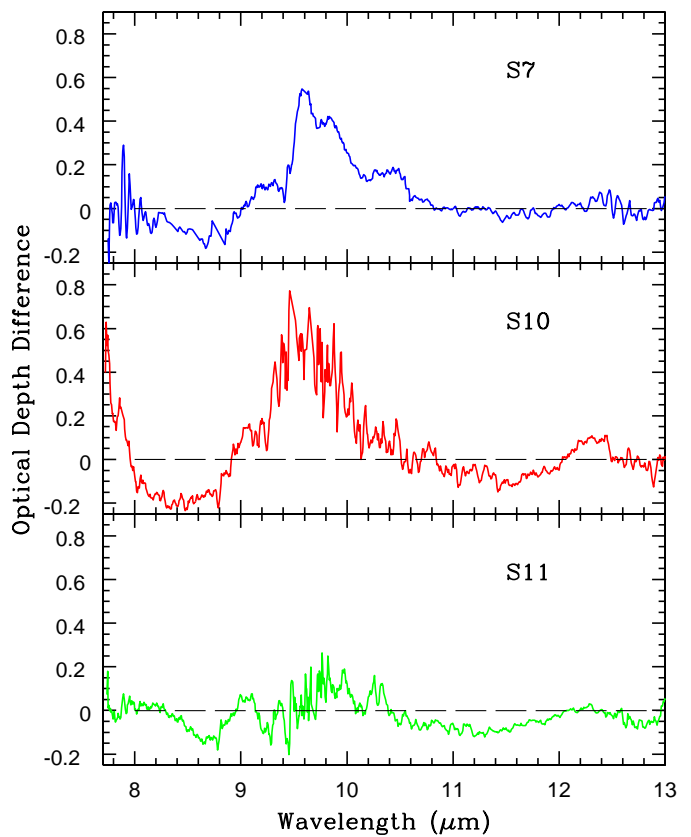}}
\caption{
Difference spectra of the silicate optical depth profiles between the observed and Galactic centre (GC) of the sources S7, S10 and S11.  
}
\label{sil_diff}
\end {figure}

We compare our silicate profiles to that of dense ISM in the line-of-sight towards Galactic centre from 
\citet{2006ApJ...637..774C}. This comparison is shown in Fig.~\ref{sil_norm}. Our normalisation at 11~\mic 
shows that the profiles match well for wavelengths longer than 10.4~\mic and shorter than 8.7~$\mu$m.  To visualise this better, 
we have plotted the difference spectra with respect to the Galactic centre  in Fig.~\ref{sil_diff} following \citet{2011A&A...526A.152V}. The differences are relatively large 
for S7 and S10. S11 shows smaller differences indicating that the dust composition is similar to that of Galactic centre. For S7, the signal-to-noise ratio is larger and the difference spectrum clearly 
shows substructure in the silicate profiles. There are two peaks, one at 9.6~\mic and the other at 9.8~$\mu$m. This could be 
either due to the combination of silicates of different composition along the line-of-sight, i.e. interstellar and circumstellar silicates or due to ice features. 
\citet{2010ApJ...718.1100B} find substructures within the silicates feature (8-13~$\mu$m) that they attribute to ices such as NH$_3$ at 9~\mic and 
CH$_3$OH at 9.7~$\mu$m. In fact, a prominent water ice absorption feature at 3.4~$\mu$m has been detected towards this region by \cite{2002AJ....124.2790I}. We also see two 
plateaus at 40\% level around 10.4~\mic and 9.3~\mic in the difference spectrum of S7.   The plateau near 9.3~\mic is also seen for S10 and S11.
This implies that there is more absorption towards S7 and S10 between 9.3 and 10.4~\mic when compared to ISM. On the other hand, this difference 
is not so marked in the case of S11. The profile of S11 matches remarkably 
well with the Galactic centre profile. For S10, there is absorption at wavelengths shorter than 8~\mic that is not seen 
towards S7 and S11 (Fig.~\ref{mir_spec}). A search for prominent ice features detected at wavelengths close to but shorter than 7.5~\mic 
shows that there is a feature at 6.0~\mic (mostly due to amorphous water ice), an unidentified feature at 6.9~\mic 
\citep{2005ApJ...635L.145K} usually attributed to NH$_4^+$ \citep{2003A&A...398.1049S} and a 7.7~\mic CH$_4$ ice feature. 
The shorter wavelength absorption (i.e. $\lambda\sim7.7$~$\mu$m) in S10 is at least 20\% that of silicate feature while the 7.7~\mic CH$_4$ ice feature 
is usually very weak \citep{2008ApJ...678..985B}, indicating that this absorption in S11 is likely to be the red edge of the  6.9~\mic absorption feature seen in the 
spectra of several protostars, including massive ones \citep{2001A&A...376..254K}. Thus, the silicate features imply that the cluster members are 
objects deeply embedded, and ices are an important constituent here.

We also note that the mid-infrared SED of S11 shows a decrease at wavelengths longer than 12.5~$\mu$m (see Fig.~\ref{mir_spec}). This could be because (i) we are seeing the 
blue edge of the silicate 18~\mic broad absorption feature, or (ii) because of a decrease in continuum fluxes. We are inclined to think it 
is the former because photometry shows that the flux is lowest at 17.7~\mic as compared to 12~\mic and 19~\mic (Table~\ref{fluxes}), 
with lower likelihood of a decreasing spectral energy distribution in mid-infrared. For S10, the photometric fluxes are increasing in wavelength and silicate absorption if present at 18~\mic is likely to be weak. 
For S7, we do not have photometric fluxes close to 18~\mic to be able to comment on the nature of the 18~\mic silicate absorption.
The consequence of the above is that, the dust  composition and distribution is different along the 
lines-of-sight of S11, as compared to S10 and S7 (as both of these show similar difference-spectra for silicate profiles at 10~\mic discussed earlier). An interesting implication (if we assume similar silicate composition) would be that the radiation environment modifies the shape of the silicate features in embedded 
early evolutionary stages. This is discussed in the next section where we discuss the evolutionary stages of cluster members on the basis of other tracers. 
As the composition of silicates is unlikely to change over scales of 0.1~pc, we speculate that the nature of silicates 
(fraction of amorphous / crystalline content), presence of ices along the line-of-sight, coagulation leading to porosity could be a few reasons for the 
differing silicates profiles. It is possible that these objects are associated with disks and the differing silicate profiles could be due to variations in the disk inclination along the line-of-sight, since it is known that the grain properties in disks undergo modifications \citep{2009ASPC..414...99M}.

\section{Evolutionary stage of Cluster members}

IRAS~18511 has been investigated spectroscopically in near 
and mid-infrared in order to understand the nature the cluster members. Of the 9 sources examined in 
near-infrared, the six objects that have high probability of belonging to the cluster are S7, S8, S10, S11, AS1 and AS2 (as discussed in Sect. 3.2). 
Among these, the three brightest at mid-infrared  (S7, S10, S11) have been examined further using mid-infrared 
imaging and spectroscopy. The rest of the cluster members (S8, AS1 and AS2) with rising near-infrared SEDs are faint in mid-infrared and are 
likely to be lower mass young stellar objects. The luminosity of this region is driven by S7, S10, and S11. These three belong to the same cloud/clump as the long wavelength (sub)millimetre images (such as JCMT) show a single clump with emission peaking at the location of S7 and a shoulder of emission towards S10 and S11. 

\begin {figure}
\resizebox{\hsize}{!}{\includegraphics[height=9.0cm,angle=-90]{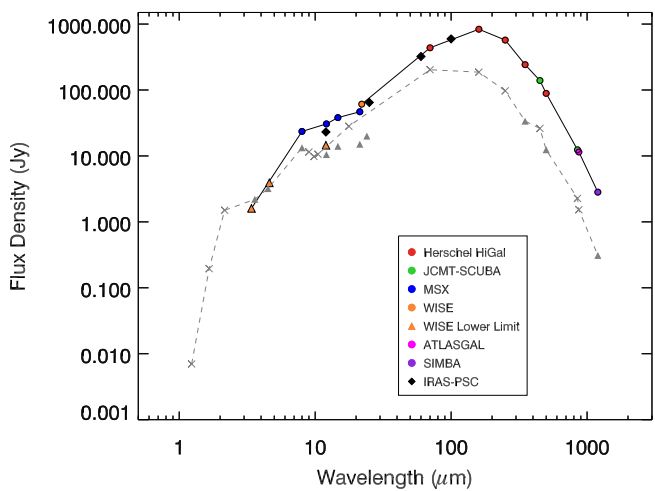}}

\caption{Spectral energy distribution of the IRAS 18511 region using flux densities in the aperture shown in Fig.~\ref{higal}.
The flux densities from MSX, WISE, \textit{Herschel} Hi-GAL, JCMT-SCUBA, ATLASGAL and SEST-SIMBA are shown using solid circles.
The triangles represent lower limits and diamonds represent IRAS-Point Source Catalog fluxes. The crosses and grey triangles represent flux densities and lower limits to flux densities of S7, respectively. In addition to the fluxes from the above mentioned images used for estimating the bolometric luminosity of S7, 2MASS and Spitzer-IRAC flux densities and lower limits have been used. The grey dashed line represents the SED under which the area is integrated to obtain a limit to the bolometric luminosity of S7. 
}
\label{sed}
\end {figure}

We have constructed the spectral energy distribution of the cloud associated with IRAS~18511 within a region of size
$1.6\times1.2$~pc$^2$ around the peak emission. The aperture used for estimating the SED is shown in 
Fig.~\ref{higal}.
As mentioned earlier, this region is selected 
such that the effect of filamentary structures and clumps in neighbouring regions are excluded. The SED is constructed 
using a number of wavelength bands: Wide-field Infrared Survey Explorer \citep[WISE,][]{2010AJ....140.1868W}, 
\textit{Herschel} Hi-GAL, MSX, JCMT-SCUBA, ATLASGAL, and SEST-SIMBA. 
The flux values from the IRAS-Point Source Catalog are also shown for reference. 
The SED is shown in Fig.~\ref{sed} and the bolometric luminosity obtained by integrating under the SED is $1.8\times10^4$~L$_\odot$.  

We have estimated a limit to the bolometric luminosity of S7 by integrating the flux density within a circular region of radius $10''$ centred on S7. While the 2MASS and VLT-VISIR observations give fairly accurate flux densities of S7, the circular region is considered specifically for estimating flux densities in the images with low resolution. We have considered the flux densities as lower limits at those wavelengths where the beamsize is larger than the region used for integrating the flux densities ($20''$). The SED of S7 is shown in Fig.~\ref{sed} in grey. By integrating under the dashed line shown in the figure, we obtain $\mathrm{L_{bol}(S7)} \gtrsim 10^4\,\, \mathrm{L_\odot}$.

The young cluster associated with IRAS~18511 lies along an unusual line-of-sight. The dust and ice absorption features towards 
the bright cluster members show relatively large optical depths. This is  evident 
not only from the silicate absorption feature presented here but also from the investigation of 3.4~\mic absorption feature 
characteristic of hydrocarbons present in the diffuse interstellar medium \citep{2012A&A...537A..27G}. The total luminosity of this region  
is $1.8\times10^4$~L$_\odot$, the total mass of the cloud is $1140\pm20$~M$_\odot$, and dust temperature in the range $19 - 26$~K. 
The bolometric luminosity of the entire region is consistent with a single B0.5 - B0 type ZAMS star \citep{1984ApJ...283..165T}. 
Below, we discuss the properties of a few cluster members in detail.

\subsection{S7}

\begin {figure}
\resizebox{\hsize}{!}{\includegraphics[height=9.0cm,angle=-90]{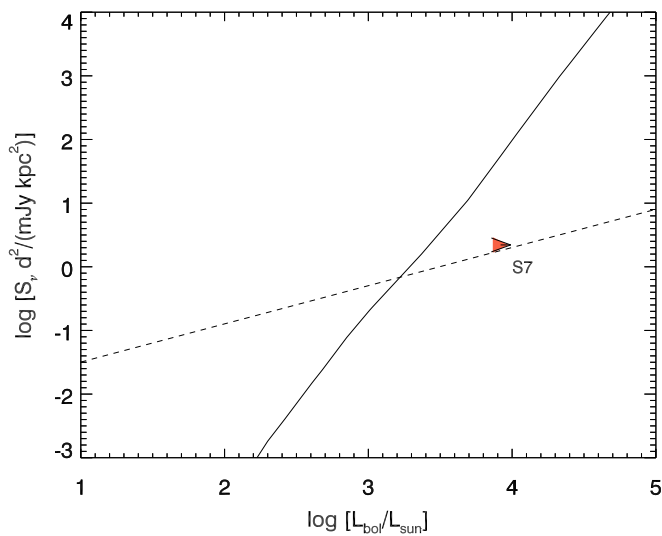}}

\caption{Plot of radio luminosity at 8 GHz versus bolometric luminosity. The solid line represents the radio luminosity expected from photoionization while the dashed line represents the fit by Anglada et al. (2015) based on observed radio jets over a wide bolometric luminosity range of $1-10^5$~L$_\odot$. The direction of arrow head indicates that the bolometric luminosity of S7 is a lower limit. 
}
\label{radio_bol}
\end {figure}

S7 is the brightest member of the cluster across the entire wavelength regime investigated (i.e near-infrared to millimetre). 
Unlike the {\it Spitzer}-IRAC images (VTW07), high resolution mid-infrared images show the presence of a single object whose ZAMS spectral type is likely to be earlier than $\sim$B0.5 based on the bolometric luminosity estimate. Weak radio emission with a flux density of $0.18\pm0.03$~mJy has been detected towards S7 at 8~GHz using sensitive EVLA measurements  by  \citet{2011ApJ...739L...9S}. This is the sole radio detection of S7 as it has not been detected at other frequencies by \citet{2011ApJ...739L...9S} and VTW07. At 23 GHz, the radio emission upper limit is 0.35 mJy, implying a radio spectral 
index $\alpha<0.87$ between 8 and 23 GHz \citep{2011ApJ...739L...9S}. In the absence of a specific spectral index value, we can only speculate on the possibilities giving rise to large spectral indices. Apart from optically thick emission from spherical and homogeneous \hii~regions where $\alpha\sim2$ \citep{1967ApJ...147..471M}, hypercompact \hii~regions are believed to have steep spectral indices, $\alpha\sim+1$ \citep{2004ApJS..154..553S}. In addition, the spectral index of a spherical envelope around a hot star with an electron density distribution of the form $\mathrm{n_e}\propto r^{-2}$ (due to stellar winds, for example) is found to be 0.6 \citep{1975A&A....39....1P}. More recently, \citet{2016ApJ...818...52T} have estimated the radio spectral index to be between 0.4 and 0.7 for outflow confined \hii~regions based on the core accretion model. 
An approach to distinguish between the jet/wind sources and \hii~regions would be to use the correlation between the radio and bolometric luminosity \citep{2015aska.confE.121A}.  \citet{2016ApJ...818...52T} have used the 8~GHz radio luminosity to show that the ultracompact and hypercompact \hii~regions lie in the vicinity of the Lyman continuum limit, segregated from sources classifed as massive young stellar objects with winds (their Fig.~18). We have plotted S7 in a similar figure shown in Fig.~\ref{radio_bol} where S7 is shown as a triangle with the direction of arrow-head indicating that the bolometric luminosity estimate is likely a lower limit. An implicit assumption here is that the flux density at 8~GHz is optically thin. The radio luminosity for various ZAMS spectral types at 8~GHz is shown as a solid line with the Lyman continuum flux values taken from \citet{1984ApJ...283..165T}. The dashed line in  Fig.~\ref{radio_bol} represents the best fit corresponding to observed radio jet sources by  \citet{2015aska.confE.121A} as

\begin{equation}
\left(\frac{S_\nu d^2}{\rm mJy\,\,kpc^2}\right) = 0.008 \left( \mathrm{\frac{L_{bol}}{L_\odot}}\right)^{0.6} 
\end{equation}

Here, $S_\nu$ is the radio flux density at 8~GHz, $d$ is the distance to the source and $\rm{L_{bol}}$ is the bolometric luminosity. S7 with a luminosity larger than $10^4$~L$_\odot$ lies in the region occupied by wind sources that are described as massive young stellar objects with ionised stellar winds. The ionised material in these radio weak massive objects  is believed to originate from winds in the star-disk system itself and these are likely precursors to hypercompact \hii~regions \citep{2007prpl.conf..181H}. That S7 is a likely wind source finds corroboration in the asymmetric profile of the Br-$\gamma$ line in Fig.~\ref{brgam} with a larger red wing compared to the blue edge (${\rm v_{red}}\sim800$~km/s). 

The luminosity in the Br-$\gamma$ line is 5.2~\Lsol if we consider de-reddening of the flux density by $\rm{A_K}\sim3.9$~mag \citep{1985ApJ...288..618R}
for  $\rm{A_V}\sim35$~mag derived from the silicate absorption feature. 
The Br-$\gamma$ emission could arise from the ionisation in a star-disk interaction region (magnetospheric accretion or outflows) seen in low mass young stellar objects such as T-Tauri stars \citep{1996ApJ...456..292N, 1990ApJ...356..646N} as well as intermediate mass young stellar objects \citep{2008A&A...489.1157K}, 
or from the ionisation of the surrounding molecular cloud. We are of the view that S7 is in the former phase based on considerations discussed earlier.
We can get an order of magnitude estimate of the accretion luminosity if we assume mass accretion to be responsible for the Br-$\gamma$ emission. The relation obtained by  \citet{2004AJ....128.1294C} is as follows:

\begin{equation}
\log_{10} \left(\frac{\rm{L_{acc}}}{\rm{L_\odot}}\right) = 0.9\times\left[\log_{10}\left(\frac{\rm{L_{Br\gamma}}}{\rm{L_\odot}}\right) + 4\right] - 0.7 
\end{equation}

Here $\rm{L_{acc}}$ and $\rm{L_{Br-\gamma}}$ represent the accretion and Br-$\gamma$ luminosities, respectively. This relation although proposed for low mass young stellar objects has been extended to intermediate mass Herbig Ae/Be objects \citep{2006A&A...459..837G} and applied to massive young stellar objects  as well \citep{2013MNRAS.430.1125C}. Applying Eqn. (8) to $\rm{L_{Br-\gamma}}$ estimated earlier, we obtain the  accretion luminosity  as $L_{\rm acc}\sim3500$~\Lsol for S7. This is close to one-third of the bolometric luminosity derived by other considerations above and is consistent with high accretion luminosities seen towards massive objects in the RMS survey \citep{2013MNRAS.430.1125C}. 

We had carried out high resolution mid-infrared spectroscopy in order to search for the presence of [\ion{Ne}{ii}] but did not detect this line.  [\ion{Ne}{ii}] is a fine-structure 
line excited by ultraviolet photons of energy $>21.6$ eV and detection of this line traces hot gas very close to the young massive star, generally seen in compact and ultracompact \hii~regions \citep{2005ApJ...631..381Z, 2003ApJ...596.1053J} and also sighted towards few high mass protostellar candidates \citep{2008ApJ...673..954C}. 
In the case of S7, we believe that the high circumstellar density, extinction due to the molecular cloud and spectral type of exciting star could be the reasons for non-detection. The peak column density determined from the column density map towards S7 is $7.3\times10^{22}$~cm$^{-2}$. This corresponds to a core density of $\mathrm{n_H}\sim\mathrm{n_e}\sim 4\times10^5$~cm$^{-3}$ assuming that the column density along the line-of-sight is associated with the core that is the size of a pixel ($6''\sim0.1$~pc). This is of the order of the critical density  
$\mathrm{n_{crit}}\sim3.6\times10^5$~cm$^{-3}$ required for collisional de-excitation of [\ion{Ne}{ii}] by electrons \citep{2000ApJ...541..779T}. The density is expected to be even higher close to the central object leading to de-excitation of ions by collisions. In our case, the lack of detection could also be attributed to optical depth effects. Contrary to one's naive expectation, extinction could play a role in the non-detection of this mid-infrared line towards dense cores, for example towards the \hii~region W33-B2 \citep{1998AJ....115.2504B}. Finally, S7 being a luminous infrared object with little radio emission, it is possible that very little of ionised neon is present here. 

Both methanol as well as water masers have been detected towards S7. The 44 GHz Class I methanol maser has been detected by \citet{2004ApJS..155..149K} and more recently by \citet{2016ApJS..222...18G} close to S7. Both the measurements are interferometric leading to high positional accuracy (better than the beam size of $2''\times$1\farcs5). The 6.7 GHz Class II methanol maser has also been detected here by  \citet{2010A&A...517A..56F}. The radiatively excited Class II masers are associated with hot molecular cores, UC\hii~regions and near-IR sources associated with massive young stellar objects, while the collisionally pumped Class I maser sources are generally offset and believed to form at the interfaces between outflows and ambient material or expansion of \hii~regions. The presence of Class I methanol maser is supported by the CO measurements towards S7 that indicate the presence of an outflow through the broad, non-gaussian profile with line wings.  However, due to the presence of alternate velocity components along the line-of-sight, it is difficult to discern the outflow through mapping \citep{2005ApJ...625..864Z}.  Three water maser spots (velocity components) have been detected towards this region \citep{2011MNRAS.418.1689U}. OH masers that are known to be associated
with an advanced stage of the appearance of UC\hii~region have not been detected here by \citet{2007A&A...465..865E}. However, these authors detected OH maser emission in a position offset from the IRAS position ($>2\arcmin$) towards the north-west, implying that it is unlikely to be associated with the 18511 molecular cloud. 
According to the sequence proposed by \citet{2010MNRAS.401.2219B}, the presence of water masers, both the classes of methanol masers, and absence of OH maser suggests that from the maser viewpoint, S7 is in relatively intermediate evolutionary stage, prior to or in early UC\hii~region phase. Our results discussed earlier point towards  the former, i.e. a phase prior to the hypercompact and ultrcompact~\hii~region phase.

The mid-infrared spectrum comprises an optically thick silicate absorption profile and ice-features implying a cold dense environment. The lack of ionic emission lines and 
PAH features is another strong indication of the youth of S7. Based on an estimate of the bolometric luminosity, we assert that S7 is the most 
massive object in the young protocluster. Its radio nature would ascertain that it is in a phase where it is still 
accreting mass and therefore likely to become more massive before the accretion stops \citep{2007prpl.conf..181H}. It can, therefore be categorised as a high mass protostellar object \citep{1998AJ....115.2504B}. The properties of S7 are compiled in Table~\ref{comprop}.

\begin{table}
\scriptsize
\caption{Comparison of properties of cluster members: S7, S10 and S11.}
\begin{tabular}{l c c c} \hline \hline
Property & S7 & S10 & S11 \\ \hline
 & & &\\
H$_2$O maser$^a$ & Y & ... & ... \\
CH$_3$OH Class I Maser$^b$ & Y & N & N    \\
CH$_3$OH Class II Maser$^c$ & Y & ... & ...    \\
OH Maser$^d$ & N & N & N    \\
Radio detection$^e$ & Y ($\alpha<0.87$) & N & Y ($\alpha\sim0.58\pm0.41$) \\
Bol. Luminosity (L$_\odot$) & $\gtrsim10^4$ & ... & $\sim6000^f$    \\
\hline
\end{tabular}
\vskip 0.2cm
Y - Detection; N - Non-detection; ... - No data available \\
$^a$ \citet{2011MNRAS.418.1689U}, Beamsize $\sim30''$\\
$^b$ \citet{2004ApJS..155..149K}, \citet{2016ApJS..222...18G},  Synthesized Beam $\sim$2\farcs0$\times$1\farcs4\\
$^c$ \citet{2010A&A...517A..56F}, Beamsize $\sim120''$\\
$^d$ \citet{2007A&A...465..865E}, Beamsize $\sim3.5'\times19'$\\
$^e$ $\alpha-$ Spectral Index \citep{2011ApJ...739L...9S} \\
$^f$ Based on assumption of optically thin radio emission from a spherical ionised gas distribution.
\label{comprop}
\end{table}

\subsection{S11}

The projected distance between S11 and S7 is 0.3~pc. The molecular CO emission towards IRAS~18511 has two peaks. The brightest 
peak is at S11 while the secondary peak is at S7. No methanol Class I maser emission detected here although 
radio emission is detected at 8.5, 15 and 23 GHz \citep{1999ApJS..125..143W, 2011ApJ...739L...9S}. The radio emission flux density ranges from 0.65 - 1.15 mJy. 
It is possible that the emission is diffuse and not point-like as we see diffuse emission 
at mid-infrared wavelengths that we associate with a photodissociated region (PDR, discussed below). The present resolution of radio images does not allow us to determine whether there is any ionised gas associated with $\rm{S11}'$. If we assume the radio flux density to be associated with S11 alone, then we estimate the Lyman continuum photon flux to be $2.3\times10^{45}$~s$^{-1}$ assuming that the ionised gas emission is optically thin at 23.1~GHz. This corresponds to a luminosity of $\sim6\times10^3$~\Lsol and a spectral type of B1 for S11 \citep{1984ApJ...283..165T}. In the absence of an independent measure of bolometric luminosity, it is not possible to plot this object in Fig.~\ref{radio_bol} and estimate whether the emission is from the associated \hii~region or jet/wind source. The spectral index value $0.58\pm0.41$  indicates that it is likely to be a jet/wind source. However, the error in spectral index value is also large. If it is indeed a ionised jet or wind source,  the bolometric luminosity of S11 would be considerably higher  ($\sim10^5$~L$_\odot$ if we apply Eqn. 7) than the value estimated earlier. In addition, we do not see any hint of the wind in the Br-$\gamma$ profile (Fig.~\ref{brgam}), unlike S7. These considerations incline us to regard S11 as a relatively evolved object, probably in the ultracompact \hii~region phase. 

Close to S11, we have detected diffuse emission along with the knot $\rm{S11}'$. The diffuse mid-infrared emission associated with young stellar objects could be elongated or otherwise. It has been 
shown that such extended emission could be due to outflow cavities, circumstellar disks  or accretion envelopes 
\citep{2006ApJ...642L..57D}. Extended mid-infrared emission 
around massive young stellar objects have been detected, in many cases within the \hii~regions indicating that hot dust can be present 
within these regions \citep{2009ApJ...698..488M}. In our case, $\rm{S11}'$ being bright in PAH bands implies that we are probably probing the 
the PDR near the edge of the \hii~region 
\citep{2012ApJ...749L..21S}. As the extended emission is seen to the west and north-west of S11 and not in other directions, this suggests that density inhomgeneities in the vicinity of S11 is the likely cause. One cannot rule out scattered light from a cavity created by an outflow from S11 providing a glimpse of 
regions closer to S11. Mid-infrared emission from cavities created by jets have been observed towards several 
massive protostars \citep{2007IAUS..237..407D} and the PDRs at the edges of the ionised cavities have been attributed to PAH emission from bipolar outflows \citep{2009MNRAS.393..354P}. Alternately, it is possible that $\rm{S11}'$ corresponds to a distinct faint source in the vicinity of S11.  In the literature, we find that PAH emission has been detected towards young stellar objects with disks: a substantial fraction of Herbig Ae/Be stars \citep{2004A&A...426..151A} and few T-Tauri stars \citep{2006A&A...459..545G}. However, these objects suffer low extinction and the fraction of PAH detection from embedded objects is very low \citep{2009A&A...495..837G, 2000IAUS..197...97V}. Compact PAH emission has also been imaged towards a few young objects by \citet{2000IAUS..197...97V}. As the cluster members of IRAS~18511 are embedded, the former possibility of $\rm{S11}'$ being associated with a cavity appears more encouraging. More sensitive observations in mid-infrared along with high resolution molecular line observations can possibly help confirm this.   Although S11 has the largest optical depth in silicate 
(present work), ices and hydrocarbon features \citep{2012A&A...537A..27G}, the silicate feature of S11 
is closer to that of Galactic centre and local diffuse ISM.  The large optical depth can be explained by the fact that S11 is more embedded than S7 (possibly through clumpy medium). A large extinction could also be explained by the viewing angle in presence of a disk like structure if we assume that $\rm{S11}'$ is indeed created by an outflow from S11.  Considering that (i) diffuse emission is seen towards S11 unlike S7, (ii)Class II methanol maser is not detected towards S11 while it is detected towards S7, (iii) S11 is probably in the \hii~region phase, and (iii) S11 has lower bolometric luminosity  
than S7, we speculate that S11 is less massive but more as evolved compared to S7.   Thus, the protostellar object likely to become the most massive star is less evolved as compared to a lower mass star. It is possible that the evolution of S7 was initiated subsequent to that of S11. On the other hand, if S7 and S11 are co-eval, then this result has a bearing on the evolutionary  mechanism of massive stars. Simulations have shown that the evolution of a massive protostar cannot be treated as an independent problem and should also involve the evolution of the host molecular core, leading to variation in properties such as the mass accretion rate \citep{2013ApJ...772...61K}.

\subsection{S10}

S10 is located to the south-east of S7 at a projected distance of nearly 0.3~pc. The shape of the silicate absorption feature is similar 
to that of S7 although the optical depth is larger. This object although bright in mid-infrared has no associated radio emission. Being in close proximity to S11 (projected distance $\sim0.1$~pc), it is difficult to ascertain its bolometric luminosity. 
From the luminosity considerations, it would appear that S10 has lower bolometric luminosity as compared to S7 and S11 (see Table~\ref{comprop}). We therefore anticipate that S10 is of lower mass than S7 or S11. Alternately, there is a possibility that S10 is a low mass and low luminosity object on its way to becoming a massive protostar. However, unlike S7 no Class I methanol maser is detected here.

The silicate feature of S10 is closer in appearance to that of S7 as compared to S11.
This is clear from the comparison of wavelength at which the absorption peaks and the width of the silicate feature (Table~\ref{sil_tab}), 
that manifest in the difference spectra shown in Fig.~\ref{sil_diff}. It is difficult to speculate on the nature of this similarity in view of the limited information we have of this object. \\

\noindent
A census of silicate absorption features towards 
hundreds of massive young stellar objects in the Large Magellanic Cloud shows that the 
evolutionary stage of the young stellar objects can be categorised based on  silicate features, emission lines and PAHs \citep{2009ApJ...699..150S}. Evidences point towards S11 being relatively more evolved as compared to the S7 and we find a variation in their silicate profiles where the silicate feature of the former is wider and peaks at a longer wavelength compared to the latter.
Our study of massive young objects from the same cluster, although younger and more embedded than the sample considered by \citet{2009ApJ...699..150S} show that silicate absorption profiles can be combined with other multiwavelength tracers to assess the evolutionary stage of the embedded object. However, further studies of silicate features as well as other tracers from the cluster members in 
different evolutionary stages are required to confirm this.

\section{Summary}

We have examined the cluster members and distribution of cold gas in the region associated with IRAS~18511+0146. Near-infrared spectroscopy
of seven objects show a rising SED. Of these, the three brightest in mid-infrared (S7, S10, S11) have been investigated through spectroscopy and 
show silicate absorption features with large optical depths. S7 is brightest at near, mid and far-infrared as well as at millimetre wavelengths and is the 
most luminous source in this region. The temperature and density distributions peak close to S7 indicating active star formation here. S11 is associated 
with diffuse emission in mid-infrared wavelengths which is bright in PAH bands and could be interpreted as the photodissociated region in the vicinity of S11. 
S7 is speculated to be younger and more massive than S11 based on its luminosity,  radio emission, and presence of methanol masers. 
The shape of the silicate feature towards S11 is different from S7 and S10 in terms of the peak absorption wavelength and the width of feature. S7 and S10 show substructures 
in their silicate absorption feature that could be due to ices or the nature of silicates as the composition is unlikely to be different across the molecular cloud. The 
bolometric luminosity of IRAS~18511+0146 is estimated to be $1.8\times10^4$~L$_\odot$, equivalent to a single B0.5-B0 ZAMS star.

\begin{acknowledgements}
We would like to thank A. G. G. M. Tielens for providing us details of their silicate profiles and A. S{\'a}nchez-Monge for letting us use  the radio images. We are grateful to F. Massi for  assistance in reduction of the near-infrared 
spectra. We also thank L. Christensen (VLT), A. Ederoclite and
 C. Snodgrass (NTT) for the support and friendly atmosphere provided while
carrying out the observations. CMW acknowledges travel support from Science Foundation,
 Ireland (Grant 13/ERC/I12907).
\end{acknowledgements}

\bibliographystyle{aa}
\bibliography{ref_svig}

\begin{thebibliography}{105}
\expandafter\ifx\csname natexlab\endcsname\relax\def\natexlab#1{#1}\fi

\bibitem[{{Acke} \& {van den Ancker}(2004)}]{2004A&A...426..151A}
{Acke}, B. \& {van den Ancker}, M.~E. 2004, \aap, 426, 151

\bibitem[{{Andr{\'e}} {et~al.}(2007){Andr{\'e}}, {Belloche}, {Motte}, \&
  {Peretto}}]{2007A&A...472..519A}
{Andr{\'e}}, P., {Belloche}, A., {Motte}, F., \& {Peretto}, N. 2007, \aap, 472,
  519

\bibitem[{{Andr{\'e}} {et~al.}(2010){Andr{\'e}}, {Men'shchikov}, {Bontemps},
  {K{\"o}nyves}, {Motte}, {Schneider}, {Didelon}, {Minier}, {Saraceno},
  {Ward-Thompson}, {di Francesco}, {White}, {Molinari}, {Testi}, {Abergel},
  {Griffin}, {Henning}, {Royer}, {Mer{\'{\i}}n}, {Vavrek}, {Attard},
  {Arzoumanian}, {Wilson}, {Ade}, {Aussel}, {Baluteau}, {Benedettini},
  {Bernard}, {Blommaert}, {Cambr{\'e}sy}, {Cox}, {di Giorgio}, {Hargrave},
  {Hennemann}, {Huang}, {Kirk}, {Krause}, {Launhardt}, {Leeks}, {Le Pennec},
  {Li}, {Martin}, {Maury}, {Olofsson}, {Omont}, {Peretto}, {Pezzuto}, {Prusti},
  {Roussel}, {Russeil}, {Sauvage}, {Sibthorpe}, {Sicilia-Aguilar}, {Spinoglio},
  {Waelkens}, {Woodcraft}, \& {Zavagno}}]{2010A&A...518L.102A}
{Andr{\'e}}, P., {Men'shchikov}, A., {Bontemps}, S., {et~al.} 2010, \aap, 518,
  L102

\bibitem[{{Anglada} {et~al.}(2015){Anglada}, {Rodr{\'{\i}}guez}, \&
  {Carrasco-Gonzalez}}]{2015aska.confE.121A}
{Anglada}, G., {Rodr{\'{\i}}guez}, L.~F., \& {Carrasco-Gonzalez}, C. 2015,
  Advancing Astrophysics with the Square Kilometre Array (AASKA14), 121

\bibitem[{{Aniano} {et~al.}(2011){Aniano}, {Draine}, {Gordon}, \&
  {Sandstrom}}]{2011PASP..123.1218A}
{Aniano}, G., {Draine}, B.~T., {Gordon}, K.~D., \& {Sandstrom}, K. 2011, \pasp,
  123, 1218

\bibitem[{{Beck} {et~al.}(1998){Beck}, {Kelly}, \&
  {Lacy}}]{1998AJ....115.2504B}
{Beck}, S.~C., {Kelly}, D.~M., \& {Lacy}, J.~H. 1998, \aj, 115, 2504

\bibitem[{{Beckwith} {et~al.}(1990){Beckwith}, {Sargent}, {Chini}, \&
  {Guesten}}]{1990AJ.....99..924B}
{Beckwith}, S.~V.~W., {Sargent}, A.~I., {Chini}, R.~S., \& {Guesten}, R. 1990,
  \aj, 99, 924

\bibitem[{{Beltr{\'a}n} {et~al.}(2006){Beltr{\'a}n}, {Brand}, {Cesaroni},
  {Fontani}, {Pezzuto}, {Testi}, \& {Molinari}}]{2006A&A...447..221B}
{Beltr{\'a}n}, M.~T., {Brand}, J., {Cesaroni}, R., {et~al.} 2006, \aap, 447,
  221

\bibitem[{{Boogert} {et~al.}(2008){Boogert}, {Pontoppidan}, {Knez}, {Lahuis},
  {Kessler-Silacci}, {van Dishoeck}, {Blake}, {Augereau}, {Bisschop},
  {Bottinelli}, {Brooke}, {Brown}, {Crapsi}, {Evans}, {Fraser}, {Geers},
  {Huard}, {J{\o}rgensen}, {{\"O}berg}, {Allen}, {Harvey}, {Koerner}, {Mundy},
  {Padgett}, {Sargent}, \& {Stapelfeldt}}]{2008ApJ...678..985B}
{Boogert}, A.~C.~A., {Pontoppidan}, K.~M., {Knez}, C., {et~al.} 2008, \apj,
  678, 985

\bibitem[{{Bottinelli} {et~al.}(2010){Bottinelli}, {Boogert}, {Bouwman},
  {Beckwith}, {van Dishoeck}, {{\"O}berg}, {Pontoppidan}, {Linnartz}, {Blake},
  {Evans}, \& {Lahuis}}]{2010ApJ...718.1100B}
{Bottinelli}, S., {Boogert}, A.~C.~A., {Bouwman}, J., {et~al.} 2010, \apj, 718,
  1100

\bibitem[{{Bowey} {et~al.}(2003){Bowey}, {Adamson}, \&
  {Yates}}]{2003MNRAS.340.1173B}
{Bowey}, J.~E., {Adamson}, A.~J., \& {Yates}, J.~A. 2003, \mnras, 340, 1173

\bibitem[{{Breen} {et~al.}(2010){Breen}, {Ellingsen}, {Caswell}, \&
  {Lewis}}]{2010MNRAS.401.2219B}
{Breen}, S.~L., {Ellingsen}, S.~P., {Caswell}, J.~L., \& {Lewis}, B.~E. 2010,
  \mnras, 401, 2219

\bibitem[{{Calvet} {et~al.}(2004){Calvet}, {Muzerolle}, {Brice{\~n}o},
  {Hern{\'a}ndez}, {Hartmann}, {Saucedo}, \& {Gordon}}]{2004AJ....128.1294C}
{Calvet}, N., {Muzerolle}, J., {Brice{\~n}o}, C., {et~al.} 2004, \aj, 128, 1294

\bibitem[{{Campbell} {et~al.}(2008){Campbell}, {Sridharan}, {Beuther}, {Lacy},
  {Hora}, {Zhu}, {Kassis}, {Saito}, {De Buizer}, {Fung}, \&
  {Johnson}}]{2008ApJ...673..954C}
{Campbell}, M.~F., {Sridharan}, T.~K., {Beuther}, H., {et~al.} 2008, \apj, 673,
  954

\bibitem[{{Chiar} {et~al.}(2007){Chiar}, {Ennico}, {Pendleton}, {Boogert},
  {Greene}, {Knez}, {Lada}, {Roellig}, {Tielens}, {Werner}, \&
  {Whittet}}]{2007ApJ...666L..73C}
{Chiar}, J.~E., {Ennico}, K., {Pendleton}, Y.~J., {et~al.} 2007, \apjl, 666,
  L73

\bibitem[{{Chiar} \& {Tielens}(2006)}]{2006ApJ...637..774C}
{Chiar}, J.~E. \& {Tielens}, A.~G.~G.~M. 2006, \apj, 637, 774

\bibitem[{{Comer{\'o}n} {et~al.}(2004){Comer{\'o}n}, {Torra}, {Chiappini},
  {Figueras}, {Ivanov}, \& {Ribas}}]{2004A&A...425..489C}
{Comer{\'o}n}, F., {Torra}, J., {Chiappini}, C., {et~al.} 2004, \aap, 425, 489

\bibitem[{{Connelley} \& {Greene}(2010)}]{2010AJ....140.1214C}
{Connelley}, M.~S. \& {Greene}, T.~P. 2010, \aj, 140, 1214

\bibitem[{{Cooper} {et~al.}(2013){Cooper}, {Lumsden}, {Oudmaijer}, {Hoare},
  {Clarke}, {Urquhart}, {Mottram}, {Moore}, \& {Davies}}]{2013MNRAS.430.1125C}
{Cooper}, H.~D.~B., {Lumsden}, S.~L., {Oudmaijer}, R.~D., {et~al.} 2013,
  \mnras, 430, 1125

\bibitem[{{Csengeri} {et~al.}(2014){Csengeri}, {Urquhart}, {Schuller}, {Motte},
  {Bontemps}, {Wyrowski}, {Menten}, {Bronfman}, {Beuther}, {Henning}, {Testi},
  {Zavagno}, \& {Walmsley}}]{2014A&A...565A..75C}
{Csengeri}, T., {Urquhart}, J.~S., {Schuller}, F., {et~al.} 2014, \aap, 565,
  A75

\bibitem[{{Cyganowski} {et~al.}(2007){Cyganowski}, {Brogan}, \&
  {Hunter}}]{2007AJ....134..346C}
{Cyganowski}, C.~J., {Brogan}, C.~L., \& {Hunter}, T.~R. 2007, \aj, 134, 346

\bibitem[{{De Buizer}(2006)}]{2006ApJ...642L..57D}
{De Buizer}, J.~M. 2006, \apjl, 642, L57

\bibitem[{{De Buizer}(2007)}]{2007IAUS..237..407D}
{De Buizer}, J.~M. 2007, in IAU Symposium, Vol. 237, IAU Symposium, ed. B.~G.
  {Elmegreen} \& J.~{Palous}, 407--407

\bibitem[{{Demyk} {et~al.}(1999){Demyk}, {Jones}, {Dartois}, {Cox}, \&
  {D'Hendecourt}}]{1999A&A...349..267D}
{Demyk}, K., {Jones}, A.~P., {Dartois}, E., {Cox}, P., \& {D'Hendecourt}, L.
  1999, \aap, 349, 267

\bibitem[{{Dupac} {et~al.}(2003){Dupac}, {Bernard}, {Boudet}, {Giard},
  {Lamarre}, {M{\'e}ny}, {Pajot}, {Ristorcelli}, {Serra}, {Stepnik}, \&
  {Torre}}]{2003A&A...404L..11D}
{Dupac}, X., {Bernard}, J.-P., {Boudet}, N., {et~al.} 2003, \aap, 404, L11

\bibitem[{{Edris} {et~al.}(2007){Edris}, {Fuller}, \&
  {Cohen}}]{2007A&A...465..865E}
{Edris}, K.~A., {Fuller}, G.~A., \& {Cohen}, R.~J. 2007, \aap, 465, 865

\bibitem[{{Faimali} {et~al.}(2012){Faimali}, {Thompson}, {Hindson}, {Urquhart},
  {Pestalozzi}, {Carey}, {Shenoy}, {Veneziani}, {Molinari}, \&
  {Clark}}]{2012MNRAS.426..402F}
{Faimali}, A., {Thompson}, M.~A., {Hindson}, L., {et~al.} 2012, \mnras, 426,
  402

\bibitem[{{Fontani} {et~al.}(2010){Fontani}, {Cesaroni}, \&
  {Furuya}}]{2010A&A...517A..56F}
{Fontani}, F., {Cesaroni}, R., \& {Furuya}, R.~S. 2010, \aap, 517, A56

\bibitem[{{F{\"o}rster Schreiber}(2000)}]{2000AJ....120.2089F}
{F{\"o}rster Schreiber}, N.~M. 2000, \aj, 120, 2089

\bibitem[{{Garcia Lopez} {et~al.}(2006){Garcia Lopez}, {Natta}, {Testi}, \&
  {Habart}}]{2006A&A...459..837G}
{Garcia Lopez}, R., {Natta}, A., {Testi}, L., \& {Habart}, E. 2006, \aap, 459,
  837

\bibitem[{{Geers} {et~al.}(2006){Geers}, {Augereau}, {Pontoppidan},
  {Dullemond}, {Visser}, {Kessler-Silacci}, {Evans}, {van Dishoeck}, {Blake},
  {Boogert}, {Brown}, {Lahuis}, \& {Mer{\'{\i}}n}}]{2006A&A...459..545G}
{Geers}, V.~C., {Augereau}, J.-C., {Pontoppidan}, K.~M., {et~al.} 2006, \aap,
  459, 545

\bibitem[{{Geers} {et~al.}(2009){Geers}, {van Dishoeck}, {Pontoppidan},
  {Lahuis}, {Crapsi}, {Dullemond}, \& {Blake}}]{2009A&A...495..837G}
{Geers}, V.~C., {van Dishoeck}, E.~F., {Pontoppidan}, K.~M., {et~al.} 2009,
  \aap, 495, 837

\bibitem[{{Godard} {et~al.}(2012){Godard}, {Geballe}, {Dartois}, \& {Mu{\~n}oz
  Caro}}]{2012A&A...537A..27G}
{Godard}, M., {Geballe}, T.~R., {Dartois}, E., \& {Mu{\~n}oz Caro}, G.~M. 2012,
  \aap, 537, A27

\bibitem[{{G{\'o}mez-Ruiz} {et~al.}(2016){G{\'o}mez-Ruiz}, {Kurtz}, {Araya},
  {Hofner}, \& {Loinard}}]{2016ApJS..222...18G}
{G{\'o}mez-Ruiz}, A.~I., {Kurtz}, S.~E., {Araya}, E.~D., {Hofner}, P., \&
  {Loinard}, L. 2016, \apjs, 222, 18

\bibitem[{{Greene} \& {Lada}(1996)}]{1996AJ....112.2184G}
{Greene}, T.~P. \& {Lada}, C.~J. 1996, \aj, 112, 2184

\bibitem[{{Greene} \& {Lada}(2000)}]{2000AJ....120..430G}
{Greene}, T.~P. \& {Lada}, C.~J. 2000, \aj, 120, 430

\bibitem[{{Griffin} {et~al.}(2010){Griffin}, {Abergel}, {Abreu}, {Ade},
  {Andr{\'e}}, {Augueres}, {Babbedge}, {Bae}, {Baillie}, {Baluteau}, {Barlow},
  {Bendo}, {Benielli}, {Bock}, {Bonhomme}, {Brisbin}, {Brockley-Blatt},
  {Caldwell}, {Cara}, {Castro-Rodriguez}, {Cerulli}, {Chanial}, {Chen},
  {Clark}, {Clements}, {Clerc}, {Coker}, {Communal}, {Conversi}, {Cox},
  {Crumb}, {Cunningham}, {Daly}, {Davis}, {de Antoni}, {Delderfield}, {Devin},
  {di Giorgio}, {Didschuns}, {Dohlen}, {Donati}, {Dowell}, {Dowell}, {Duband},
  {Dumaye}, {Emery}, {Ferlet}, {Ferrand}, {Fontignie}, {Fox}, {Franceschini},
  {Frerking}, {Fulton}, {Garcia}, {Gastaud}, {Gear}, {Glenn}, {Goizel},
  {Griffin}, {Grundy}, {Guest}, {Guillemet}, {Hargrave}, {Harwit}, {Hastings},
  {Hatziminaoglou}, {Herman}, {Hinde}, {Hristov}, {Huang}, {Imhof}, {Isaak},
  {Israelsson}, {Ivison}, {Jennings}, {Kiernan}, {King}, {Lange}, {Latter},
  {Laurent}, {Laurent}, {Leeks}, {Lellouch}, {Levenson}, {Li}, {Li},
  {Lilienthal}, {Lim}, {Liu}, {Lu}, {Madden}, {Mainetti}, {Marliani}, {McKay},
  {Mercier}, {Molinari}, {Morris}, {Moseley}, {Mulder}, {Mur}, {Naylor},
  {Nguyen}, {O'Halloran}, {Oliver}, {Olofsson}, {Olofsson}, {Orfei}, {Page},
  {Pain}, {Panuzzo}, {Papageorgiou}, {Parks}, {Parr-Burman}, {Pearce},
  {Pearson}, {P{\'e}rez-Fournon}, {Pinsard}, {Pisano}, {Podosek}, {Pohlen},
  {Polehampton}, {Pouliquen}, {Rigopoulou}, {Rizzo}, {Roseboom}, {Roussel},
  {Rowan-Robinson}, {Rownd}, {Saraceno}, {Sauvage}, {Savage}, {Savini},
  {Sawyer}, {Scharmberg}, {Schmitt}, {Schneider}, {Schulz}, {Schwartz},
  {Shafer}, {Shupe}, {Sibthorpe}, {Sidher}, {Smith}, {Smith}, {Smith},
  {Spencer}, {Stobie}, {Sudiwala}, {Sukhatme}, {Surace}, {Stevens}, {Swinyard},
  {Trichas}, {Tourette}, {Triou}, {Tseng}, {Tucker}, {Turner}, {Vaccari},
  {Valtchanov}, {Vigroux}, {Virique}, {Voellmer}, {Walker}, {Ward}, {Waskett},
  {Weilert}, {Wesson}, {White}, {Whitehouse}, {Wilson}, {Winter}, {Woodcraft},
  {Wright}, {Xu}, {Zavagno}, {Zemcov}, {Zhang}, \&
  {Zonca}}]{2010A&A...518L...3G}
{Griffin}, M.~J., {Abergel}, A., {Abreu}, A., {et~al.} 2010, \aap, 518, L3

\bibitem[{{Hambly} {et~al.}(2008){Hambly}, {Collins}, {Cross}, {Mann}, {Read},
  {Sutorius}, {Bond}, {Bryant}, {Emerson}, {Lawrence}, {Rimoldini}, {Stewart},
  {Williams}, {Adamson}, {Hirst}, {Dye}, \& {Warren}}]{2008MNRAS.384..637H}
{Hambly}, N.~C., {Collins}, R.~S., {Cross}, N.~J.~G., {et~al.} 2008, \mnras,
  384, 637

\bibitem[{{Hildebrand}(1983)}]{1983QJRAS..24..267H}
{Hildebrand}, R.~H. 1983, \qjras, 24, 267

\bibitem[{{Hoare} {et~al.}(2007){Hoare}, {Kurtz}, {Lizano}, {Keto}, \&
  {Hofner}}]{2007prpl.conf..181H}
{Hoare}, M.~G., {Kurtz}, S.~E., {Lizano}, S., {Keto}, E., \& {Hofner}, P. 2007,
  Protostars and Planets V, 181

\bibitem[{{Ishii} {et~al.}(2002){Ishii}, {Nagata}, {Chrysostomou}, \&
  {Hough}}]{2002AJ....124.2790I}
{Ishii}, M., {Nagata}, T., {Chrysostomou}, A., \& {Hough}, J.~H. 2002, \aj,
  124, 2790

\bibitem[{{Jaeger} {et~al.}(1994){Jaeger}, {Mutschke}, {Begemann}, {Dorschner},
  \& {Henning}}]{1994A&A...292..641J}
{Jaeger}, C., {Mutschke}, H., {Begemann}, B., {Dorschner}, J., \& {Henning}, T.
  1994, \aap, 292, 641

\bibitem[{{Jaffe} {et~al.}(2003){Jaffe}, {Zhu}, {Lacy}, \&
  {Richter}}]{2003ApJ...596.1053J}
{Jaffe}, D.~T., {Zhu}, Q., {Lacy}, J.~H., \& {Richter}, M. 2003, \apj, 596,
  1053

\bibitem[{{Keane} {et~al.}(2001){Keane}, {Tielens}, {Boogert}, {Schutte}, \&
  {Whittet}}]{2001A&A...376..254K}
{Keane}, J.~V., {Tielens}, A.~G.~G.~M., {Boogert}, A.~C.~A., {Schutte}, W.~A.,
  \& {Whittet}, D.~C.~B. 2001, \aap, 376, 254

\bibitem[{{Kemper} {et~al.}(2004){Kemper}, {Vriend}, \&
  {Tielens}}]{2004ApJ...609..826K}
{Kemper}, F., {Vriend}, W.~J., \& {Tielens}, A.~G.~G.~M. 2004, \apj, 609, 826

\bibitem[{{Kessler-Silacci} {et~al.}(2005){Kessler-Silacci}, {Hillenbrand},
  {Blake}, \& {Meyer}}]{2005ApJ...622..404K}
{Kessler-Silacci}, J.~E., {Hillenbrand}, L.~A., {Blake}, G.~A., \& {Meyer},
  M.~R. 2005, \apj, 622, 404

\bibitem[{{Klein} {et~al.}(2005){Klein}, {Posselt}, {Schreyer}, {Forbrich}, \&
  {Henning}}]{2005ApJS..161..361K}
{Klein}, R., {Posselt}, B., {Schreyer}, K., {Forbrich}, J., \& {Henning}, T.
  2005, \apjs, 161, 361

\bibitem[{{Knez} {et~al.}(2005){Knez}, {Boogert}, {Pontoppidan},
  {Kessler-Silacci}, {van Dishoeck}, {Evans}, {Augereau}, {Blake}, \&
  {Lahuis}}]{2005ApJ...635L.145K}
{Knez}, C., {Boogert}, A.~C.~A., {Pontoppidan}, K.~M., {et~al.} 2005, \apjl,
  635, L145

\bibitem[{{Kraus}(2009)}]{2009A&A...494..253K}
{Kraus}, M. 2009, \aap, 494, 253

\bibitem[{{Kraus} {et~al.}(2008){Kraus}, {Hofmann}, {Benisty}, {Berger},
  {Chesneau}, {Isella}, {Malbet}, {Meilland}, {Nardetto}, {Natta}, {Preibisch},
  {Schertl}, {Smith}, {Stee}, {Tatulli}, {Testi}, \&
  {Weigelt}}]{2008A&A...489.1157K}
{Kraus}, S., {Hofmann}, K.-H., {Benisty}, M., {et~al.} 2008, \aap, 489, 1157

\bibitem[{{Kuiper} \& {Yorke}(2013)}]{2013ApJ...772...61K}
{Kuiper}, R. \& {Yorke}, H.~W. 2013, \apj, 772, 61

\bibitem[{{Kurtz} {et~al.}(2004){Kurtz}, {Hofner}, \&
  {{\'A}lvarez}}]{2004ApJS..155..149K}
{Kurtz}, S., {Hofner}, P., \& {{\'A}lvarez}, C.~V. 2004, \apjs, 155, 149

\bibitem[{{Lada} \& {Lada}(2003)}]{2003ARA&A..41...57L}
{Lada}, C.~J. \& {Lada}, E.~A. 2003, \araa, 41, 57

\bibitem[{{Lagage} {et~al.}(2004){Lagage}, {Pel}, {Authier}, {Belorgey},
  {Claret}, {Doucet}, {Dubreuil}, {Durand}, {Elswijk}, {Girardot}, {K{\"a}ufl},
  {Kroes}, {Lortholary}, {Lussignol}, {Marchesi}, {Pantin}, {Peletier},
  {Pirard}, {Pragt}, {Rio}, {Schoenmaker}, {Siebenmorgen}, {Silber}, {Smette},
  {Sterzik}, \& {Veyssiere}}]{2004Msngr.117...12L}
{Lagage}, P.~O., {Pel}, J.~W., {Authier}, M., {et~al.} 2004, The Messenger,
  117, 12

\bibitem[{{Launhardt} {et~al.}(2013){Launhardt}, {Stutz}, {Schmiedeke},
  {Henning}, {Krause}, {Balog}, {Beuther}, {Birkmann}, {Hennemann},
  {Kainulainen}, {Khanzadyan}, {Linz}, {Lippok}, {Nielbock}, {Pitann}, {Ragan},
  {Risacher}, {Schmalzl}, {Shirley}, {Stecklum}, {Steinacker}, \&
  {Tackenberg}}]{2013A&A...551A..98L}
{Launhardt}, R., {Stutz}, A.~M., {Schmiedeke}, A., {et~al.} 2013, \aap, 551,
  A98

\bibitem[{{Lawrence} {et~al.}(2007){Lawrence}, {Warren}, {Almaini}, {Edge},
  {Hambly}, {Jameson}, {Lucas}, {Casali}, {Adamson}, {Dye}, {Emerson},
  {Foucaud}, {Hewett}, {Hirst}, {Hodgkin}, {Irwin}, {Lodieu}, {McMahon},
  {Simpson}, {Smail}, {Mortlock}, \& {Folger}}]{2007MNRAS.379.1599L}
{Lawrence}, A., {Warren}, S.~J., {Almaini}, O., {et~al.} 2007, \mnras, 379,
  1599

\bibitem[{{Lucas} {et~al.}(2008){Lucas}, {Hoare}, {Longmore}, {Schr{\"o}der},
  {Davis}, {Adamson}, {Bandyopadhyay}, {de Grijs}, {Smith}, {Gosling},
  {Mitchison}, {G{\'a}sp{\'a}r}, {Coe}, {Tamura}, {Parker}, {Irwin}, {Hambly},
  {Bryant}, {Collins}, {Cross}, {Evans}, {Gonzalez-Solares}, {Hodgkin},
  {Lewis}, {Read}, {Riello}, {Sutorius}, {Lawrence}, {Drew}, {Dye}, \&
  {Thompson}}]{2008MNRAS.391..136L}
{Lucas}, P.~W., {Hoare}, M.~G., {Longmore}, A., {et~al.} 2008, \mnras, 391, 136

\bibitem[{{Lumsden} {et~al.}(2013){Lumsden}, {Hoare}, {Urquhart}, {Oudmaijer},
  {Davies}, {Mottram}, {Cooper}, \& {Moore}}]{2013ApJS..208...11L}
{Lumsden}, S.~L., {Hoare}, M.~G., {Urquhart}, J.~S., {et~al.} 2013, \apjs, 208,
  11

\bibitem[{{Mathis}(1998)}]{1998ApJ...497..824M}
{Mathis}, J.~S. 1998, \apj, 497, 824

\bibitem[{{McKee} \& {Ostriker}(2007)}]{2007ARA&A..45..565M}
{McKee}, C.~F. \& {Ostriker}, E.~C. 2007, \araa, 45, 565

\bibitem[{{Meeus}(2009)}]{2009ASPC..414...99M}
{Meeus}, G. 2009, in Astronomical Society of the Pacific Conference Series,
  Vol. 414, Cosmic Dust - Near and Far, ed. T.~{Henning}, E.~{Gr{\"u}n}, \&
  J.~{Steinacker}, 99

\bibitem[{{Meyer} {et~al.}(1998){Meyer}, {Edwards}, {Hinkle}, \&
  {Strom}}]{1998ApJ...508..397M}
{Meyer}, M.~R., {Edwards}, S., {Hinkle}, K.~H., \& {Strom}, S.~E. 1998, \apj,
  508, 397

\bibitem[{{Mezger} \& {Henderson}(1967)}]{1967ApJ...147..471M}
{Mezger}, P.~G. \& {Henderson}, A.~P. 1967, \apj, 147, 471

\bibitem[{{Molinari} {et~al.}(2016){Molinari}, {Schisano}, {Elia},
  {Pestalozzi}, {Traficante}, {Pezzuto}, {Swinyard}, {Noriega-Crespo}, {Bally},
  {Moore}, {Plume}, {Zavagno}, {di Giorgio}, {Liu}, {Pilbratt}, {Mottram},
  {Russeil}, {Piazzo}, {Veneziani}, {Benedettini}, {Calzoletti}, {Faustini},
  {Natoli}, {Piacentini}, {Merello}, {Palmese}, {Del Grande}, {Polychroni},
  {Rygl}, {Polenta}, {Barlow}, {Bernard}, {Martin}, {Testi}, {Ali},
  {Andr{\'e}}, {Beltr{\'a}n}, {Billot}, {Carey}, {Cesaroni}, {Compi{\`e}gne},
  {Eden}, {Fukui}, {Garcia-Lario}, {Hoare}, {Huang}, {Joncas}, {Lim}, {Lord},
  {Martinavarro-Armengol}, {Motte}, {Paladini}, {Paradis}, {Peretto},
  {Robitaille}, {Schilke}, {Schneider}, {Schulz}, {Sibthorpe}, {Strafella},
  {Thompson}, {Umana}, {Ward-Thompson}, \& {Wyrowski}}]{2016A&A...591A.149M}
{Molinari}, S., {Schisano}, E., {Elia}, D., {et~al.} 2016, \aap, 591, A149

\bibitem[{{Molinari} {et~al.}(2010{\natexlab{a}}){Molinari}, {Swinyard},
  {Bally}, {Barlow}, {Bernard}, {Martin}, {Moore}, {Noriega-Crespo}, {Plume},
  {Testi}, {Zavagno}, {Abergel}, {Ali}, {Anderson}, {Andr{\'e}}, {Baluteau},
  {Battersby}, {Beltr{\'a}n}, {Benedettini}, {Billot}, {Blommaert}, {Bontemps},
  {Boulanger}, {Brand}, {Brunt}, {Burton}, {Calzoletti}, {Carey}, {Caselli},
  {Cesaroni}, {Cernicharo}, {Chakrabarti}, {Chrysostomou}, {Cohen},
  {Compiegne}, {de Bernardis}, {de Gasperis}, {di Giorgio}, {Elia}, {Faustini},
  {Flagey}, {Fukui}, {Fuller}, {Ganga}, {Garcia-Lario}, {Glenn}, {Goldsmith},
  {Griffin}, {Hoare}, {Huang}, {Ikhenaode}, {Joblin}, {Joncas}, {Juvela},
  {Kirk}, {Lagache}, {Li}, {Lim}, {Lord}, {Marengo}, {Marshall}, {Masi},
  {Massi}, {Matsuura}, {Minier}, {Miville-Desch{\^e}nes}, {Montier}, {Morgan},
  {Motte}, {Mottram}, {M{\"u}ller}, {Natoli}, {Neves}, {Olmi}, {Paladini},
  {Paradis}, {Parsons}, {Peretto}, {Pestalozzi}, {Pezzuto}, {Piacentini},
  {Piazzo}, {Polychroni}, {Pomar{\`e}s}, {Popescu}, {Reach}, {Ristorcelli},
  {Robitaille}, {Robitaille}, {Rod{\'o}n}, {Roy}, {Royer}, {Russeil},
  {Saraceno}, {Sauvage}, {Schilke}, {Schisano}, {Schneider}, {Schuller},
  {Schulz}, {Sibthorpe}, {Smith}, {Smith}, {Spinoglio}, {Stamatellos},
  {Strafella}, {Stringfellow}, {Sturm}, {Taylor}, {Thompson}, {Traficante},
  {Tuffs}, {Umana}, {Valenziano}, {Vavrek}, {Veneziani}, {Viti}, {Waelkens},
  {Ward-Thompson}, {White}, {Wilcock}, {Wyrowski}, {Yorke}, \&
  {Zhang}}]{2010A&A...518L.100M}
{Molinari}, S., {Swinyard}, B., {Bally}, J., {et~al.} 2010{\natexlab{a}}, \aap,
  518, L100

\bibitem[{{Molinari} {et~al.}(2010{\natexlab{b}}){Molinari}, {Swinyard},
  {Bally}, {Barlow}, {Bernard}, {Martin}, {Moore}, {Noriega-Crespo}, {Plume},
  {Testi}, {Zavagno}, {Abergel}, {Ali}, {Andr{\'e}}, {Baluteau}, {Benedettini},
  {Bern{\'e}}, {Billot}, {Blommaert}, {Bontemps}, {Boulanger}, {Brand},
  {Brunt}, {Burton}, {Campeggio}, {Carey}, {Caselli}, {Cesaroni}, {Cernicharo},
  {Chakrabarti}, {Chrysostomou}, {Codella}, {Cohen}, {Compiegne}, {Davis}, {de
  Bernardis}, {de Gasperis}, {Di Francesco}, {di Giorgio}, {Elia}, {Faustini},
  {Fischera}, {Fukui}, {Fuller}, {Ganga}, {Garcia-Lario}, {Giard}, {Giardino},
  {Glenn}, {Goldsmith}, {Griffin}, {Hoare}, {Huang}, {Jiang}, {Joblin},
  {Joncas}, {Juvela}, {Kirk}, {Lagache}, {Li}, {Lim}, {Lord}, {Lucas},
  {Maiolo}, {Marengo}, {Marshall}, {Masi}, {Massi}, {Matsuura}, {Meny},
  {Minier}, {Miville-Desch{\^e}nes}, {Montier}, {Motte}, {M{\"u}ller},
  {Natoli}, {Neves}, {Olmi}, {Paladini}, {Paradis}, {Pestalozzi}, {Pezzuto},
  {Piacentini}, {Pomar{\`e}s}, {Popescu}, {Reach}, {Richer}, {Ristorcelli},
  {Roy}, {Royer}, {Russeil}, {Saraceno}, {Sauvage}, {Schilke},
  {Schneider-Bontemps}, {Schuller}, {Schultz}, {Shepherd}, {Sibthorpe},
  {Smith}, {Smith}, {Spinoglio}, {Stamatellos}, {Strafella}, {Stringfellow},
  {Sturm}, {Taylor}, {Thompson}, {Tuffs}, {Umana}, {Valenziano}, {Vavrek},
  {Viti}, {Waelkens}, {Ward-Thompson}, {White}, {Wyrowski}, {Yorke}, \&
  {Zhang}}]{2010PASP..122..314M}
{Molinari}, S., {Swinyard}, B., {Bally}, J., {et~al.} 2010{\natexlab{b}},
  \pasp, 122, 314

\bibitem[{{Moorwood} {et~al.}(1998){Moorwood}, {Cuby}, \&
  {Lidman}}]{1998Msngr..91....9M}
{Moorwood}, A., {Cuby}, J.-G., \& {Lidman}, C. 1998, The Messenger, 91, 9

\bibitem[{{Morales} {et~al.}(2009){Morales}, {Mardones}, {Garay}, {Brooks}, \&
  {Pineda}}]{2009ApJ...698..488M}
{Morales}, E.~F.~E., {Mardones}, D., {Garay}, G., {Brooks}, K.~J., \& {Pineda},
  J.~E. 2009, \apj, 698, 488

\bibitem[{{Mottram} {et~al.}(2011){Mottram}, {Hoare}, {Urquhart}, {Lumsden},
  {Oudmaijer}, {Robitaille}, {Moore}, {Davies}, \&
  {Stead}}]{2011A&A...525A.149M}
{Mottram}, J.~C., {Hoare}, M.~G., {Urquhart}, J.~S., {et~al.} 2011, \aap, 525,
  A149

\bibitem[{{Najita} {et~al.}(1996){Najita}, {Carr}, \&
  {Tokunaga}}]{1996ApJ...456..292N}
{Najita}, J., {Carr}, J.~S., \& {Tokunaga}, A.~T. 1996, \apj, 456, 292

\bibitem[{{Natta} \& {Giovanardi}(1990)}]{1990ApJ...356..646N}
{Natta}, A. \& {Giovanardi}, C. 1990, \apj, 356, 646

\bibitem[{{Panagia} \& {Felli}(1975)}]{1975A&A....39....1P}
{Panagia}, N. \& {Felli}, M. 1975, \aap, 39, 1

\bibitem[{{Paradis} {et~al.}(2010){Paradis}, {Veneziani}, {Noriega-Crespo},
  {Paladini}, {Piacentini}, {Bernard}, {de Bernardis}, {Calzoletti},
  {Faustini}, {Martin}, {Masi}, {Montier}, {Natoli}, {Ristorcelli}, {Thompson},
  {Traficante}, \& {Molinari}}]{2010A&A...520L...8P}
{Paradis}, D., {Veneziani}, M., {Noriega-Crespo}, A., {et~al.} 2010, \aap, 520,
  L8

\bibitem[{{Peretto} \& {Fuller}(2009)}]{2009A&A...505..405P}
{Peretto}, N. \& {Fuller}, G.~A. 2009, \aap, 505, 405

\bibitem[{{Phillips} \& {P{\'e}rez-Grana}(2009)}]{2009MNRAS.393..354P}
{Phillips}, J.~P. \& {P{\'e}rez-Grana}, J.~A. 2009, \mnras, 393, 354

\bibitem[{{Pilbratt} {et~al.}(2010){Pilbratt}, {Riedinger}, {Passvogel},
  {Crone}, {Doyle}, {Gageur}, {Heras}, {Jewell}, {Metcalfe}, {Ott}, \&
  {Schmidt}}]{2010A&A...518L...1P}
{Pilbratt}, G.~L., {Riedinger}, J.~R., {Passvogel}, T., {et~al.} 2010, \aap,
  518, L1

\bibitem[{{Poglitsch} {et~al.}(2010){Poglitsch}, {Waelkens}, {Geis},
  {Feuchtgruber}, {Vandenbussche}, {Rodriguez}, {Krause}, {Renotte}, {van
  Hoof}, {Saraceno}, {Cepa}, {Kerschbaum}, {Agn{\`e}se}, {Ali}, {Altieri},
  {Andreani}, {Augueres}, {Balog}, {Barl}, {Bauer}, {Belbachir}, {Benedettini},
  {Billot}, {Boulade}, {Bischof}, {Blommaert}, {Callut}, {Cara}, {Cerulli},
  {Cesarsky}, {Contursi}, {Creten}, {De Meester}, {Doublier}, {Doumayrou},
  {Duband}, {Exter}, {Genzel}, {Gillis}, {Gr{\"o}zinger}, {Henning},
  {Herreros}, {Huygen}, {Inguscio}, {Jakob}, {Jamar}, {Jean}, {de Jong},
  {Katterloher}, {Kiss}, {Klaas}, {Lemke}, {Lutz}, {Madden}, {Marquet},
  {Martignac}, {Mazy}, {Merken}, {Montfort}, {Morbidelli}, {M{\"u}ller},
  {Nielbock}, {Okumura}, {Orfei}, {Ottensamer}, {Pezzuto}, {Popesso},
  {Putzeys}, {Regibo}, {Reveret}, {Royer}, {Sauvage}, {Schreiber}, {Stegmaier},
  {Schmitt}, {Schubert}, {Sturm}, {Thiel}, {Tofani}, {Vavrek}, {Wetzstein},
  {Wieprecht}, \& {Wiezorrek}}]{2010A&A...518L...2P}
{Poglitsch}, A., {Waelkens}, C., {Geis}, N., {et~al.} 2010, \aap, 518, L2

\bibitem[{{Reipurth} {et~al.}(2002){Reipurth}, {Hartmann}, {Kenyon}, {Smette},
  \& {Bouchet}}]{2002AJ....124.2194R}
{Reipurth}, B., {Hartmann}, L., {Kenyon}, S.~J., {Smette}, A., \& {Bouchet}, P.
  2002, \aj, 124, 2194

\bibitem[{{Rieke} \& {Lebofsky}(1985)}]{1985ApJ...288..618R}
{Rieke}, G.~H. \& {Lebofsky}, M.~J. 1985, \apj, 288, 618

\bibitem[{{Rodr{\'{\i}}guez-Gonz{\'a}lez}
  {et~al.}(2008){Rodr{\'{\i}}guez-Gonz{\'a}lez}, {Esquivel}, {Raga}, \&
  {Cant{\'o}}}]{2008ApJ...684.1384R}
{Rodr{\'{\i}}guez-Gonz{\'a}lez}, A., {Esquivel}, A., {Raga}, A.~C., \&
  {Cant{\'o}}, J. 2008, \apj, 684, 1384

\bibitem[{{Sadavoy} {et~al.}(2013){Sadavoy}, {Di Francesco}, {Johnstone},
  {Currie}, {Drabek}, {Hatchell}, {Nutter}, {Andr{\'e}}, {Arzoumanian},
  {Benedettini}, {Bernard}, {Duarte-Cabral}, {Fallscheer}, {Friesen},
  {Greaves}, {Hennemann}, {Hill}, {Jenness}, {K{\"o}nyves}, {Matthews},
  {Mottram}, {Pezzuto}, {Roy}, {Rygl}, {Schneider-Bontemps}, {Spinoglio},
  {Testi}, {Tothill}, {Ward-Thompson}, {White}, {JCMT}, \& {Herschel Gould Belt
  Survey Teams}}]{2013ApJ...767..126S}
{Sadavoy}, S.~I., {Di Francesco}, J., {Johnstone}, D., {et~al.} 2013, \apj,
  767, 126

\bibitem[{{Salgado} {et~al.}(2012){Salgado}, {Bern{\'e}}, {Adams}, {Herter},
  {Gull}, {Schoenwald}, {Keller}, {De Buizer}, {Vacca}, {Becklin}, {Shuping},
  {Tielens}, \& {Zinnecker}}]{2012ApJ...749L..21S}
{Salgado}, F., {Bern{\'e}}, O., {Adams}, J.~D., {et~al.} 2012, \apjl, 749, L21

\bibitem[{{S{\'a}nchez-Monge} {et~al.}(2011){S{\'a}nchez-Monge}, {Pandian}, \&
  {Kurtz}}]{2011ApJ...739L...9S}
{S{\'a}nchez-Monge}, {\'A}., {Pandian}, J.~D., \& {Kurtz}, S. 2011, \apjl, 739,
  L9

\bibitem[{{Schuller} {et~al.}(2010){Schuller}, {Beuther}, {Bontemps},
  {Bronfman}, {Carlhoff}, {Cesaroni}, {Contreras}, {Csengari}, {Deharveng},
  {Garay}, {Henning}, {Herpin}, {Immer}, {Lefloch}, {Linz}, {Mardones},
  {Menten}, {Minier}, {Molinari}, {Motte}, {Nguyen Luong}, {Nyman},
  {Rathborne}, {Reveret}, {Risacher}, {Russeil}, {Schilke}, {Schneider},
  {Tackenberg}, {Testi}, {Troost}, {Vasyunina}, {Walmsley}, {Wienen},
  {Wyrowski}, \& {Zavagno}}]{2010Msngr.141...20S}
{Schuller}, F., {Beuther}, H., {Bontemps}, S., {et~al.} 2010, The Messenger,
  141, 20

\bibitem[{{Schutte} \& {Khanna}(2003)}]{2003A&A...398.1049S}
{Schutte}, W.~A. \& {Khanna}, R.~K. 2003, \aap, 398, 1049

\bibitem[{{Seale} {et~al.}(2009){Seale}, {Looney}, {Chu}, {Gruendl}, {Brandl},
  {Chen}, {Brandner}, \& {Blake}}]{2009ApJ...699..150S}
{Seale}, J.~P., {Looney}, L.~W., {Chu}, Y.-H., {et~al.} 2009, \apj, 699, 150

\bibitem[{{Sewilo} {et~al.}(2004){Sewilo}, {Watson}, {Araya}, {Churchwell},
  {Hofner}, \& {Kurtz}}]{2004ApJS..154..553S}
{Sewilo}, M., {Watson}, C., {Araya}, E., {et~al.} 2004, \apjs, 154, 553

\bibitem[{{Takahashi} {et~al.}(2000){Takahashi}, {Matsuhara}, {Watarai}, \&
  {Matsumoto}}]{2000ApJ...541..779T}
{Takahashi}, H., {Matsuhara}, H., {Watarai}, H., \& {Matsumoto}, T. 2000, \apj,
  541, 779

\bibitem[{{Tanaka} {et~al.}(2016){Tanaka}, {Tan}, \&
  {Zhang}}]{2016ApJ...818...52T}
{Tanaka}, K.~E.~I., {Tan}, J.~C., \& {Zhang}, Y. 2016, \apj, 818, 52

\bibitem[{{Testi} {et~al.}(1998){Testi}, {Palla}, \&
  {Natta}}]{1998A&AS..133...81T}
{Testi}, L., {Palla}, F., \& {Natta}, A. 1998, \aaps, 133, 81

\bibitem[{{Thompson}(1984)}]{1984ApJ...283..165T}
{Thompson}, R.~I. 1984, \apj, 283, 165

\bibitem[{{Urquhart} {et~al.}(2011){Urquhart}, {Morgan}, {Figura}, {Moore},
  {Lumsden}, {Hoare}, {Oudmaijer}, {Mottram}, {Davies}, \&
  {Dunham}}]{2011MNRAS.418.1689U}
{Urquhart}, J.~S., {Morgan}, L.~K., {Figura}, C.~C., {et~al.} 2011, \mnras,
  418, 1689

\bibitem[{{van Breemen} {et~al.}(2011){van Breemen}, {Min}, {Chiar}, {Waters},
  {Kemper}, {Boogert}, {Cami}, {Decin}, {Knez}, {Sloan}, \&
  {Tielens}}]{2011A&A...526A.152V}
{van Breemen}, J.~M., {Min}, M., {Chiar}, J.~E., {et~al.} 2011, \aap, 526, A152

\bibitem[{{van Dishoeck} \& {van der Tak}(2000)}]{2000IAUS..197...97V}
{van Dishoeck}, E.~F. \& {van der Tak}, F.~F.~S. 2000, in IAU Symposium, Vol.
  197, From Molecular Clouds to Planetary, ed. Y.~C. {Minh} \& E.~F. {van
  Dishoeck}, 97

\bibitem[{{Vig} {et~al.}(2007){Vig}, {Testi}, {Walmsley}, {Molinari}, {Carey},
  \& {Noriega-Crespo}}]{2007A&A...470..977V}
{Vig}, S., {Testi}, L., {Walmsley}, M., {et~al.} 2007, \aap, 470, 977 (VTW07)

\bibitem[{{Wallace} \& {Hinkle}(1997)}]{1997ApJS..111..445W}
{Wallace}, L. \& {Hinkle}, K. 1997, \apjs, 111, 445

\bibitem[{{Wang} {et~al.}(2012){Wang}, {Zhang}, {Wu}, {Li}, \&
  {Zhang}}]{2012ApJ...745L..30W}
{Wang}, K., {Zhang}, Q., {Wu}, Y., {Li}, H.-b., \& {Zhang}, H. 2012, \apjl,
  745, L30

\bibitem[{{Ward-Thompson} {et~al.}(2002){Ward-Thompson}, {Andr{\'e}}, \&
  {Kirk}}]{2002MNRAS.329..257W}
{Ward-Thompson}, D., {Andr{\'e}}, P., \& {Kirk}, J.~M. 2002, \mnras, 329, 257

\bibitem[{{Ward-Thompson} {et~al.}(2010){Ward-Thompson}, {Kirk}, {Andr{\'e}},
  {Saraceno}, {Didelon}, {K{\"o}nyves}, {Schneider}, {Abergel}, {Baluteau},
  {Bernard}, {Bontemps}, {Cambr{\'e}sy}, {Cox}, {di Francesco}, {di Giorgio},
  {Griffin}, {Hargrave}, {Huang}, {Li}, {Martin}, {Men'shchikov}, {Minier},
  {Molinari}, {Motte}, {Olofsson}, {Pezzuto}, {Russeil}, {Sauvage},
  {Sibthorpe}, {Spinoglio}, {Testi}, {White}, {Wilson}, {Woodcraft}, \&
  {Zavagno}}]{2010A&A...518L..92W}
{Ward-Thompson}, D., {Kirk}, J.~M., {Andr{\'e}}, P., {et~al.} 2010, \aap, 518,
  L92

\bibitem[{{Watt} \& {Mundy}(1999)}]{1999ApJS..125..143W}
{Watt}, S. \& {Mundy}, L.~G. 1999, \apjs, 125, 143

\bibitem[{{Wolf-Chase} {et~al.}(2003){Wolf-Chase}, {Moriarty-Schieven}, {Fich},
  \& {Barsony}}]{2003MNRAS.344..809W}
{Wolf-Chase}, G., {Moriarty-Schieven}, G., {Fich}, M., \& {Barsony}, M. 2003,
  \mnras, 344, 809

\bibitem[{{Wright} {et~al.}(2010){Wright}, {Eisenhardt}, {Mainzer}, {Ressler},
  {Cutri}, {Jarrett}, {Kirkpatrick}, {Padgett}, {McMillan}, {Skrutskie},
  {Stanford}, {Cohen}, {Walker}, {Mather}, {Leisawitz}, {Gautier}, {McLean},
  {Benford}, {Lonsdale}, {Blain}, {Mendez}, {Irace}, {Duval}, {Liu}, {Royer},
  {Heinrichsen}, {Howard}, {Shannon}, {Kendall}, {Walsh}, {Larsen}, {Cardon},
  {Schick}, {Schwalm}, {Abid}, {Fabinsky}, {Naes}, \&
  {Tsai}}]{2010AJ....140.1868W}
{Wright}, E.~L., {Eisenhardt}, P.~R.~M., {Mainzer}, A.~K., {et~al.} 2010, \aj,
  140, 1868

\bibitem[{{Zhang} {et~al.}(2005){Zhang}, {Hunter}, {Brand}, {Sridharan},
  {Cesaroni}, {Molinari}, {Wang}, \& {Kramer}}]{2005ApJ...625..864Z}
{Zhang}, Q., {Hunter}, T.~R., {Brand}, J., {et~al.} 2005, \apj, 625, 864

\bibitem[{{Zhu} {et~al.}(2005){Zhu}, {Lacy}, {Jaffe}, {Richter}, \&
  {Greathouse}}]{2005ApJ...631..381Z}
{Zhu}, Q.-F., {Lacy}, J.~H., {Jaffe}, D.~T., {Richter}, M.~J., \& {Greathouse},
  T.~K. 2005, \apj, 631, 381

\bibitem[{{Zinnecker} \& {Yorke}(2007)}]{2007ARA&A..45..481Z}
{Zinnecker}, H. \& {Yorke}, H.~W. 2007, \araa, 45, 481

\end{thebibliography}

\appendix
\section{Herschel images of IRAS 18511+0146}

\begin {figure*}
\includegraphics[height=6.5cm,angle=0]{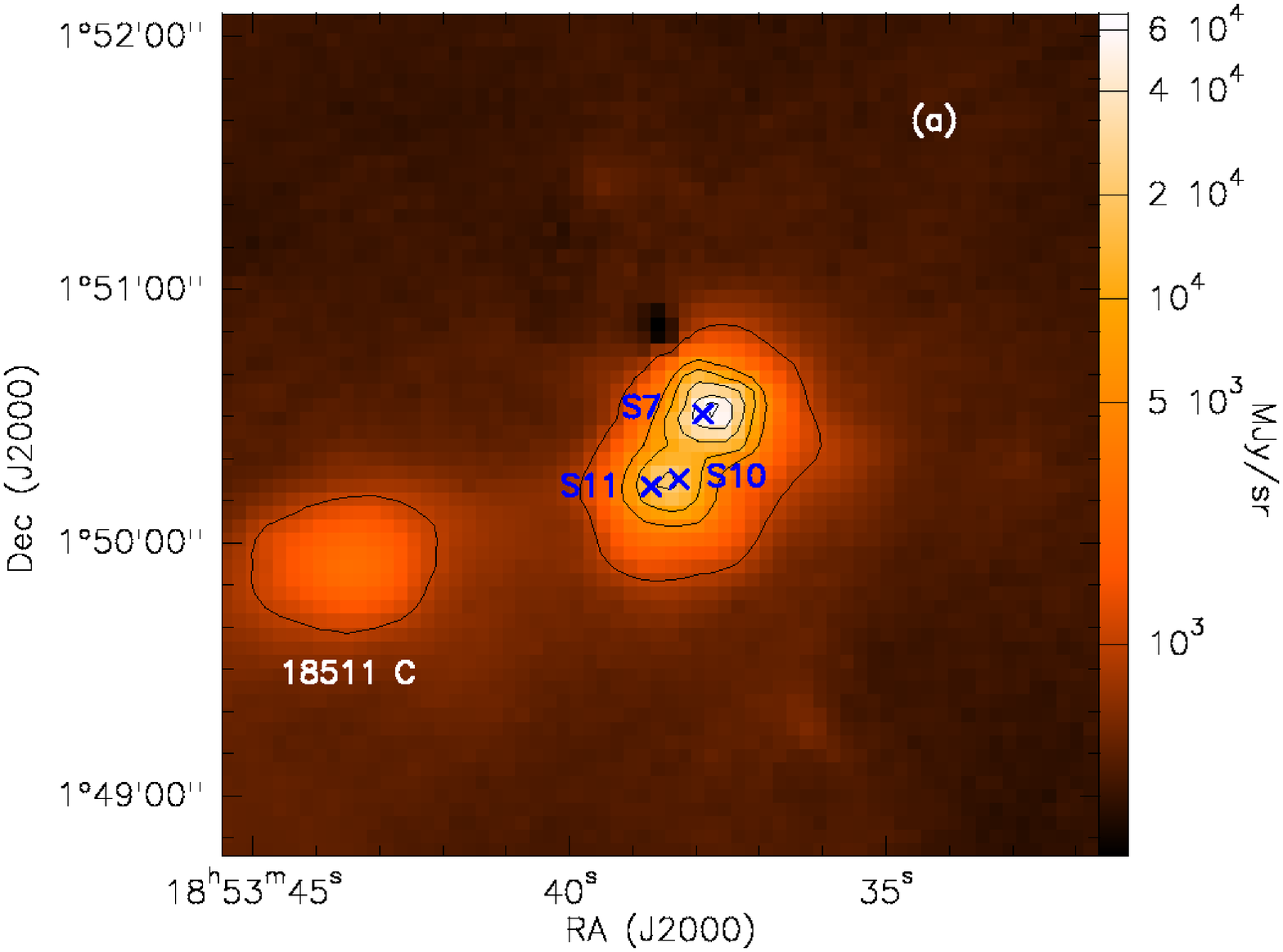} 
\includegraphics[height=6.5cm,angle=0]{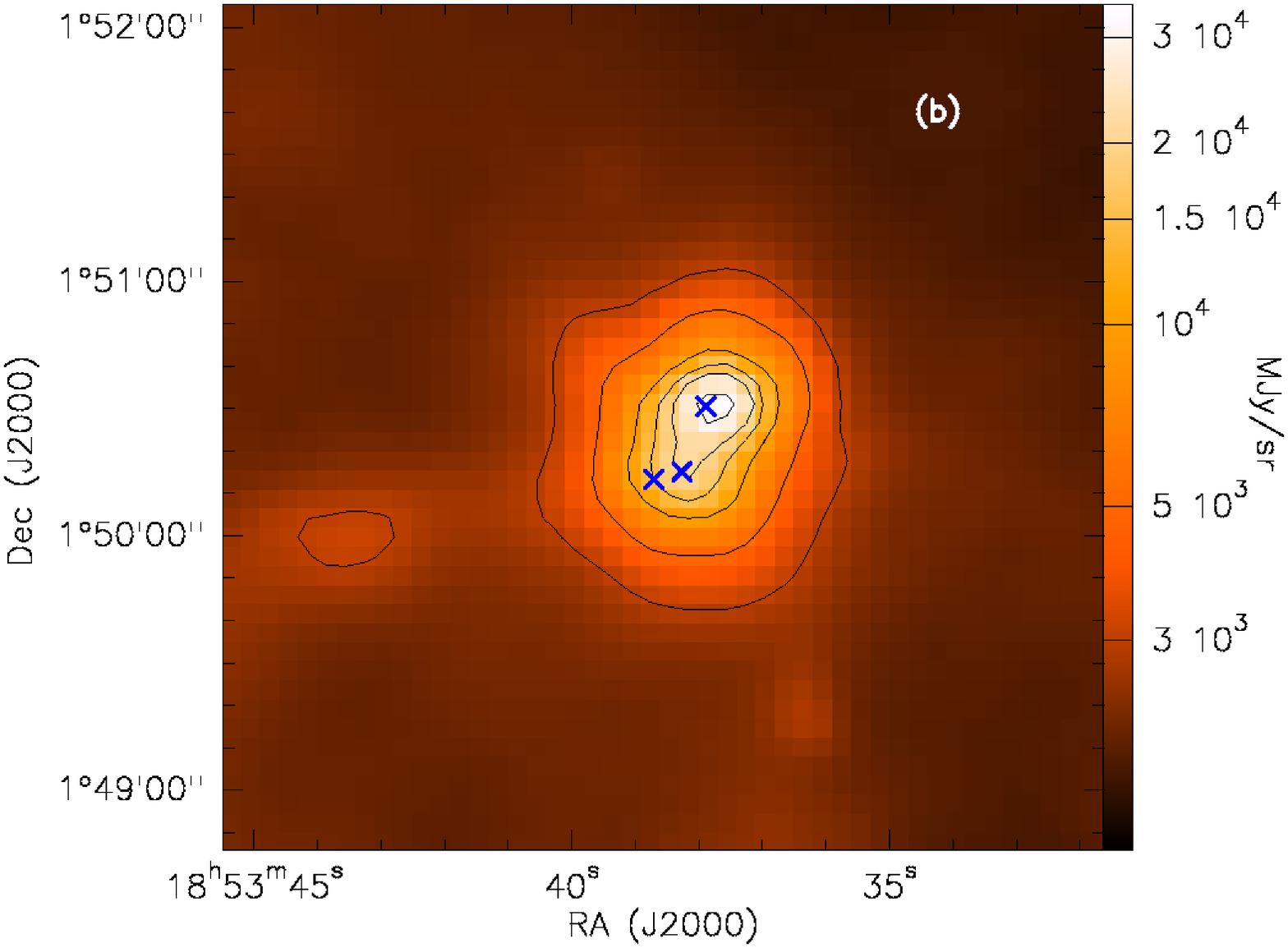}
\includegraphics[height=6.5cm,angle=0]{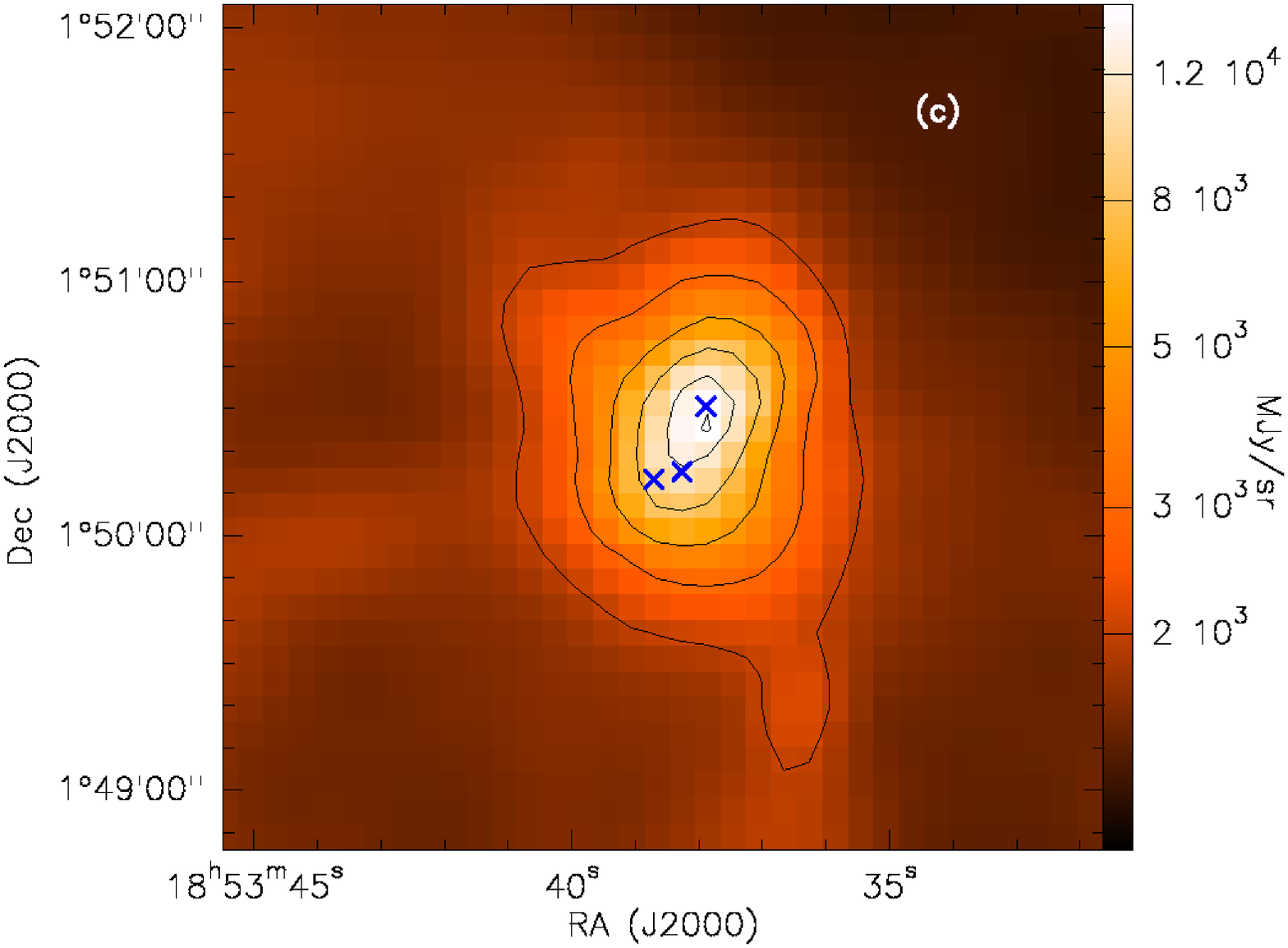}
\includegraphics[height=6.5cm,angle=0]{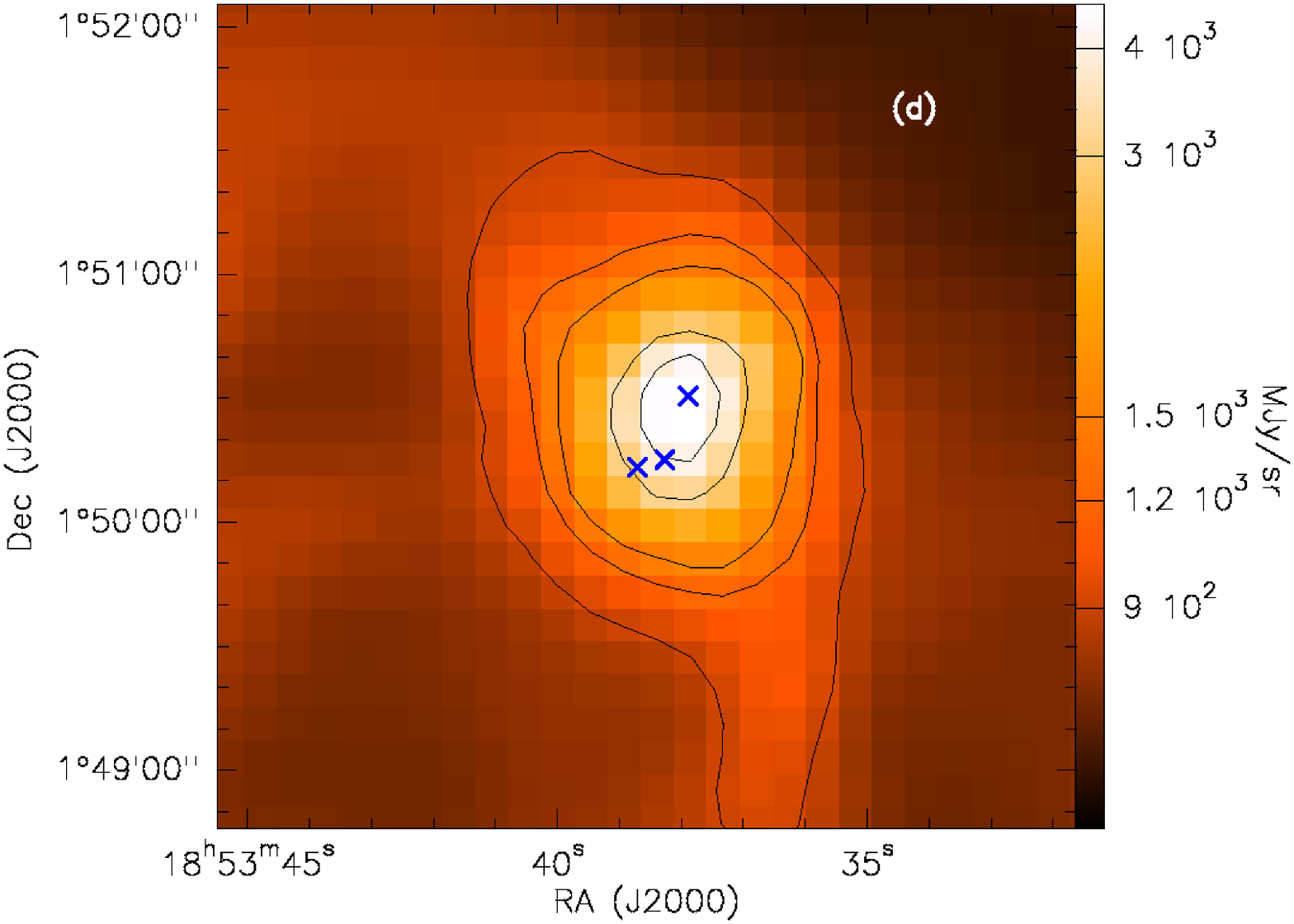} 
\includegraphics[height=6.5cm,angle=0]{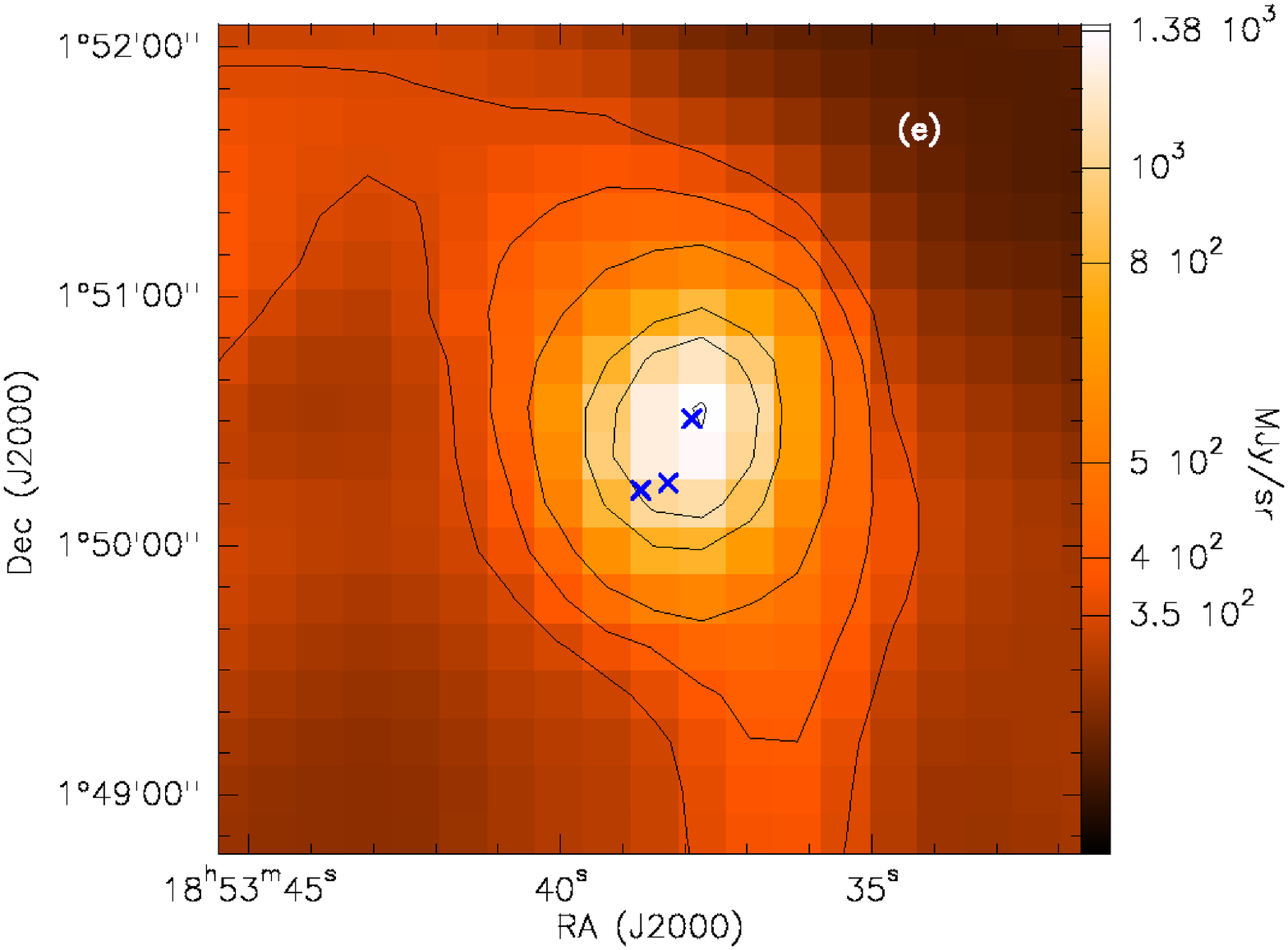}
\caption{Emission towards IRAS 18511+0146 at (a) 70~$\mu$m, (b) 160~$\mu$m, (c) 250~$\mu$m, (d) 350~$\mu$m, and (e) 500~$\mu$m using \textit{Herschel} PACs and SPIRE. The contour levels are marked on the color wedges.
}
\label{herschel}
\end {figure*}

\end{document}